%% file: 20TCAS_BEACHES_arxiv.tex
\documentclass[journal,twoside,twocolumn]{IEEEtran}
\usepackage{cite}
\usepackage[T1]{fontenc}
\usepackage{graphicx}
\usepackage{xcolor}
\usepackage{amssymb}
\usepackage{amsmath}
\usepackage{subfigure}
\usepackage{booktabs} 
\usepackage{algorithm}
\usepackage{algorithmic}
\usepackage{amsthm}
\usepackage{multirow}
\usepackage{microtype} 
\usepackage{balance}

\input{vmr-symbols-vecbold}
\input{standard-macros}

\input{defs}

\allowdisplaybreaks 
\newcommand{\yhat}[0]{\hat{\bmy}} 
\newcommand{\hhat}[0]{\hat{\bmh}} 
\newcommand{\yhate}[1]{\hat{y}_{#1}} 
\newcommand{\yshate}[1]{\hat{y}^s_{#1}} 

\newcommand{\revision}[1]{{#1}}
\newcommand{\revisionS}[1]{#1}

\safemath{\LAMA}{\textrm{LAMA}}
\safemath{\MRT}{\textrm{MRT}}
\safemath{\betamax}{\beta^\text{max}_\setO}
\safemath{\betamaxno}{\beta^\text{max}}
\safemath{\betamin}{\beta^\text{min}_\setO}
\safemath{\betaminno}{\beta^\text{min}}

\safemath{\Nomin}{\No^\textnormal{min}(\beta)}
\safemath{\Nominnobeta}{\No^\text{min}}
\safemath{\Nomax}{\No^\textnormal{max}(\beta)}
\safemath{\Nomaxnobeta}{\No^\textnormal{max}}
\safemath{\EX}{E_\textnormal{x}}
\safemath{\EXP}{\EX^\textnormal{p}}
\safemath{\Eo}{E_0}

\safemath{\tmax}{{t_\textnormal{max}}}
\safemath{\MAP}{\textrm{MAP}}
\safemath{\IO}{\textrm{IO}}
\safemath{\JO}{\textrm{JO}}
\safemath{\Nopost}{N_{0}^\textnormal{post}}
\safemath{\MT}{U}
\safemath{\MR}{B}
\safemath{\Tran}{\textnormal{T}}
\safemath{\Herm}{\textnormal{H}}
\safemath{\row}{\textnormal{r}}
\safemath{\col}{\textnormal{c}}

\safemath{\NT}{N_\textnormal{T}}
\safemath{\DSNR}{\delta \textnormal{SNR}}
\safemath{\betaMOR}{\beta^{\star}}

\newtheorem{remark}{Remark}

\begin{document}
	
\title{Beamspace Channel Estimation for Massive MIMO mmWave  Systems: Algorithm and VLSI Design}
\author{Seyed Hadi Mirfarshbafan, Alexandra Gallyas-Sanhueza,  Ramina Ghods, and Christoph Studer\thanks{S.~H.~Mirfarshbafan and A.~Gallyas-Sanhueza are with the School of Electrical and Computer Engineering at Cornell University, Ithaca, NY, and with Cornell Tech, New York, NY; email: sm2675@cornell.edu, ag753@cornell.edu}
\thanks{R.~Ghods was with the School of Electrical and Computer Engineering at Cornell University, Ithaca, NY, and is now with the School of Electrical and Computer Engineering at Carnegie Mellon University, Pittsburgh, PA; email: rghods@cs.cmu.edu}
\thanks{C.~Studer was with the School of Electrical and Computer Engineering at Cornell Tech, New York, NY, and Cornell University, Ithaca, NY, and is now with the Department of Information Technology and Electrical Engineering at ETH Zurich, Switzerland. Email: studer@ethz.ch; Web: http://iis.ee.ethz.ch}
\thanks{The work of SHM, AGS, and CS was supported in part by Xilinx, Inc.\ and by ComSenTer, one of six centers in JUMP, a Semiconductor Research Corporation (SRC) program sponsored by DARPA.
The work of RG was supported by the US National Science Foundation under grants ECCS-1408006, CCF-1535897,  CCF-1652065, CNS-1717559, and ECCS-1824379.}
\thanks{A MATLAB simulator to reproduce the results of this paper is available on GitHub: https://github.com/IIP-Group/BEACHES-simulator}\thanks{A short version of this paper has been presented at the IEEE SPAWC~\cite{ghods19}.}}

\maketitle
\input{0-abstract}
\input{1-introduction}
\input{2-system_model}
\input{3-beaches}

\input{4-implementation}

\input{5-conclusions}
\section*{Acknowledgments}
The authors thank O.~Casta\~neda for his help during the FPGA design,  C.~Jeon for discussions on SURE-based denoising, and D.~Stiverson for his preliminary exploration of low-complexity beamspace channel estimation algorithms for mmWave systems.
\appendices 
\input{6-MSEtoSURE}
\input{7-shrinkageSURE}

\input{8-SUREconvergence}

\balance
\bibliographystyle{IEEEtran}
\bibliography{bib/confs-jrnls,bib/IEEEabrv,bib/publishers,bib/vipbib}
\balance

\end{document}

%% file: vmr-symbols-vecbold.tex
\usepackage{amssymb}
\usepackage{amsfonts}
\usepackage{mathrsfs}
\usepackage{xspace}
\usepackage{bm}
\usepackage{upgreek}

\newcommand{\safemath}[2]{\newcommand{#1}{\ensuremath{#2}\xspace}}

\safemath{\bma}{\mathbf{a}}
\safemath{\bmb}{\mathbf{b}}
\safemath{\bmc}{\mathbf{c}}
\safemath{\bmd}{\mathbf{d}}
\safemath{\bme}{\mathbf{e}}
\safemath{\bmf}{\mathbf{f}}
\safemath{\bmg}{\mathbf{g}}
\safemath{\bmh}{\mathbf{h}}
\safemath{\bmi}{\mathbf{i}}
\safemath{\bmj}{\mathbf{j}}
\safemath{\bmk}{\mathbf{k}}
\safemath{\bml}{\mathbf{l}}
\safemath{\bmm}{\mathbf{m}}
\safemath{\bmn}{\mathbf{n}}
\safemath{\bmo}{\mathbf{o}}
\safemath{\bmp}{\mathbf{p}}
\safemath{\bmq}{\mathbf{q}}
\safemath{\bmr}{\mathbf{r}}
\safemath{\bms}{\mathbf{s}}
\safemath{\bmt}{\mathbf{t}}
\safemath{\bmu}{\mathbf{u}}
\safemath{\bmv}{\mathbf{v}}
\safemath{\bmw}{\mathbf{w}}
\safemath{\bmx}{\mathbf{x}}
\safemath{\bmy}{\mathbf{y}}
\safemath{\bmz}{\mathbf{z}}
\safemath{\bmzero}{\mathbf{0}}
\safemath{\bmone}{\mathbf{1}}

\bmdefine{\biad}{a}
\bmdefine{\bibd}{b}
\bmdefine{\bicd}{c}
\bmdefine{\bidd}{d}
\bmdefine{\bied}{e}
\bmdefine{\bifd}{f}
\bmdefine{\bigd}{g}
\bmdefine{\bihd}{h}
\bmdefine{\biid}{i}
\bmdefine{\bijd}{j}
\bmdefine{\bikd}{k}
\bmdefine{\bild}{l}
\bmdefine{\bimd}{m}
\bmdefine{\bind}{n}
\bmdefine{\biod}{o}
\bmdefine{\bipd}{p}
\bmdefine{\biqd}{q}
\bmdefine{\bird}{r}
\bmdefine{\bisd}{s}
\bmdefine{\bitd}{t}
\bmdefine{\biud}{u}
\bmdefine{\bivd}{v}
\bmdefine{\biwd}{w}
\bmdefine{\bixd}{x}
\bmdefine{\biyd}{y}
\bmdefine{\bizd}{z}

\bmdefine{\bixid}{\xi}
\bmdefine{\bilambdad}{\lambda}
\bmdefine{\bimud}{\mu}
\bmdefine{\bithetad}{\theta}
\bmdefine{\biphid}{\phi}
\bmdefine{\bideltad}{\delta}

\safemath{\bmia}{\biad}
\safemath{\bmib}{\bibd}
\safemath{\bmic}{\bicd}
\safemath{\bmid}{\bidd}
\safemath{\bmie}{\bied}
\safemath{\bmif}{\bifd}
\safemath{\bmig}{\bigd}
\safemath{\bmih}{\bihd}
\safemath{\bmii}{\biid}
\safemath{\bmij}{\bijd}
\safemath{\bmik}{\bikd}
\safemath{\bmil}{\bild}
\safemath{\bmim}{\bimd}
\safemath{\bmin}{\bind}
\safemath{\bmio}{\biod}
\safemath{\bmip}{\bipd}
\safemath{\bmiq}{\biqd}
\safemath{\bmir}{\bird}
\safemath{\bmis}{\bisd}
\safemath{\bmit}{\bitd}
\safemath{\bmiu}{\biud}
\safemath{\bmiv}{\bivd}
\safemath{\bmiw}{\biwd}
\safemath{\bmix}{\bixd}
\safemath{\bmiy}{\biyd}
\safemath{\bmiz}{\bizd}

\safemath{\bmxi}{\bixid}
\safemath{\bmlambda}{\bilambdad}
\safemath{\bmmu}{\bimud}
\safemath{\bmtheta}{\bithetad}
\safemath{\bmphi}{\biphid}
\safemath{\bmdelta}{\bideltad}

\safemath{\bA}{\mathbf{A}}
\safemath{\bB}{\mathbf{B}}
\safemath{\bC}{\mathbf{C}}
\safemath{\bD}{\mathbf{D}}
\safemath{\bE}{\mathbf{E}}
\safemath{\bF}{\mathbf{F}}
\safemath{\bG}{\mathbf{G}}
\safemath{\bH}{\mathbf{H}}
\safemath{\bI}{\mathbf{I}}
\safemath{\bJ}{\mathbf{J}}
\safemath{\bK}{\mathbf{K}}
\safemath{\bL}{\mathbf{L}}
\safemath{\bM}{\mathbf{M}}
\safemath{\bN}{\mathbf{N}}
\safemath{\bO}{\mathbf{O}}
\safemath{\bP}{\mathbf{P}}
\safemath{\bQ}{\mathbf{Q}}
\safemath{\bR}{\mathbf{R}}
\safemath{\bS}{\mathbf{S}}
\safemath{\bT}{\mathbf{T}}
\safemath{\bU}{\mathbf{U}}
\safemath{\bV}{\mathbf{V}}
\safemath{\bW}{\mathbf{W}}
\safemath{\bX}{\mathbf{X}}
\safemath{\bY}{\mathbf{Y}}
\safemath{\bZ}{\mathbf{Z}}

\safemath{\bZero}{\mathbf{0}}
\safemath{\bOne}{\mathbf{1}}
\safemath{\bDelta}{\mathbf{\Delta}}
\safemath{\bLambda}{\mathbf{\UpLambda}}
\safemath{\bPhi}{\mathbf{\Upphi}}
\safemath{\bSigma}{\mathbf{\Upsigma}}
\safemath{\bOmega}{\mathbf{\Upomega}}
\safemath{\bTheta}{\mathbf{\Uptheta}}

\bmdefine{\biAd}{A}
\bmdefine{\biBd}{B}
\bmdefine{\biCd}{C}
\bmdefine{\biDd}{D}
\bmdefine{\biEd}{E}
\bmdefine{\biFd}{F}
\bmdefine{\biGd}{G}
\bmdefine{\biHd}{H}
\bmdefine{\biId}{I}
\bmdefine{\biJd}{J}
\bmdefine{\biKd}{K}
\bmdefine{\biLd}{L}
\bmdefine{\biMd}{M}
\bmdefine{\biOd}{N}
\bmdefine{\biPd}{O}
\bmdefine{\biQd}{P}
\bmdefine{\biRd}{R}
\bmdefine{\biSd}{S}
\bmdefine{\biTd}{T}
\bmdefine{\biUd}{U}
\bmdefine{\biVd}{V}
\bmdefine{\biWd}{W}
\bmdefine{\biXd}{X}
\bmdefine{\biYd}{Y}
\bmdefine{\biZd}{Z}

\bmdefine{\biDelta}{\Delta}
\bmdefine{\biLambda}{\Lambda}
\bmdefine{\biPhi}{\Phi}
\bmdefine{\biSigma}{\Sigma}
\bmdefine{\biOmega}{\Omega}
\bmdefine{\biTheta}{\Theta}

\safemath{\bimA}{\biAd}
\safemath{\bimB}{\biBd}
\safemath{\bimC}{\biCd}
\safemath{\bimD}{\biDd}
\safemath{\bimE}{\biEd}
\safemath{\bimF}{\biFd}
\safemath{\bimG}{\biGd}
\safemath{\bimH}{\biHd}
\safemath{\bimI}{\biId}
\safemath{\bimJ}{\biJd}
\safemath{\bimK}{\biKd}
\safemath{\bimL}{\biLd}
\safemath{\bimM}{\biMd}
\safemath{\bimN}{\biNd}
\safemath{\bimO}{\biOd}
\safemath{\bimP}{\biPd}
\safemath{\bimQ}{\biQd}
\safemath{\bimR}{\biRd}
\safemath{\bimS}{\biSd}
\safemath{\bimT}{\biTd}
\safemath{\bimU}{\biUd}
\safemath{\bimV}{\biVd}
\safemath{\bimW}{\biWd}
\safemath{\bimX}{\biXd}
\safemath{\bimY}{\biYd}
\safemath{\bimZ}{\biZd}

\safemath{\bimDelta}{\biDelta}
\safemath{\bimLambda}{\biLambda}
\safemath{\bimPhi}{\biPhi}
\safemath{\bimSigma}{\biSigma}
\safemath{\bimOmega}{\biOmega}
\safemath{\bimTheta}{\biTheta}

\safemath{\setA}{\mathcal{A}}
\safemath{\setB}{\mathcal{B}}
\safemath{\setC}{\mathcal{C}}
\safemath{\setD}{\mathcal{D}}
\safemath{\setE}{\mathcal{E}}
\safemath{\setF}{\mathcal{F}}
\safemath{\setG}{\mathcal{G}}
\safemath{\setH}{\mathcal{H}}
\safemath{\setI}{\mathcal{I}}
\safemath{\setJ}{\mathcal{J}}
\safemath{\setK}{\mathcal{K}}
\safemath{\setL}{\mathcal{L}}
\safemath{\setM}{\mathcal{M}}
\safemath{\setN}{\mathcal{N}}
\safemath{\setO}{\mathcal{O}}
\safemath{\setP}{\mathcal{P}}
\safemath{\setQ}{\mathcal{Q}}
\safemath{\setR}{\mathcal{R}}
\safemath{\setS}{\mathcal{S}}
\safemath{\setT}{\mathcal{T}}
\safemath{\setU}{\mathcal{U}}
\safemath{\setV}{\mathcal{V}}
\safemath{\setW}{\mathcal{W}}
\safemath{\setX}{\mathcal{X}}
\safemath{\setY}{\mathcal{Y}}
\safemath{\setZ}{\mathcal{Z}}
\safemath{\emptySet}{\varnothing}

\safemath{\colA}{\mathscr{A}}
\safemath{\colB}{\mathscr{B}}
\safemath{\colC}{\mathscr{C}}
\safemath{\colD}{\mathscr{D}}
\safemath{\colE}{\mathscr{E}}
\safemath{\colF}{\mathscr{F}}
\safemath{\colG}{\mathscr{G}}
\safemath{\colH}{\mathscr{H}}
\safemath{\colI}{\mathscr{I}}
\safemath{\colJ}{\mathscr{J}}
\safemath{\colK}{\mathscr{K}}
\safemath{\colL}{\mathscr{L}}
\safemath{\colM}{\mathscr{M}}
\safemath{\colN}{\mathscr{N}}
\safemath{\colO}{\mathscr{O}}
\safemath{\colP}{\mathscr{P}}
\safemath{\colQ}{\mathscr{Q}}
\safemath{\colR}{\mathscr{R}}
\safemath{\colS}{\mathscr{S}}
\safemath{\colT}{\mathscr{T}}
\safemath{\colU}{\mathscr{U}}
\safemath{\colV}{\mathscr{V}}
\safemath{\colW}{\mathscr{W}}
\safemath{\colX}{\mathscr{X}}
\safemath{\colY}{\mathscr{Y}}
\safemath{\colZ}{\mathscr{Z}}

\safemath{\opA}{\mathbb{A}}
\safemath{\opB}{\mathbb{B}}
\safemath{\opC}{\mathbb{C}}
\safemath{\opD}{\mathbb{D}}
\safemath{\opE}{\mathbb{E}}
\safemath{\opF}{\mathbb{F}}
\safemath{\opG}{\mathbb{G}}
\safemath{\opH}{\mathbb{H}}
\safemath{\opI}{\mathbb{I}}
\safemath{\opJ}{\mathbb{J}}
\safemath{\opK}{\mathbb{K}}
\safemath{\opL}{\mathbb{L}}
\safemath{\opM}{\mathbb{M}}
\safemath{\opN}{\mathbb{N}}
\safemath{\opO}{\mathbb{O}}
\safemath{\opP}{\mathbb{P}}
\safemath{\opQ}{\mathbb{Q}}
\safemath{\opR}{\mathbb{R}}
\safemath{\opS}{\mathbb{S}}
\safemath{\opT}{\mathbb{T}}
\safemath{\opU}{\mathbb{U}}
\safemath{\opV}{\mathbb{V}}
\safemath{\opW}{\mathbb{W}}
\safemath{\opX}{\mathbb{X}}
\safemath{\opY}{\mathbb{Y}}
\safemath{\opZ}{\mathbb{Z}}
\safemath{\opZero}{\mathbb{O}}
\safemath{\identityop}{\opI}

\safemath{\veca}{\bma}
\safemath{\vecb}{\bmb}
\safemath{\vecc}{\bmc}
\safemath{\vecd}{\bmd}
\safemath{\vece}{\bme}
\safemath{\vecf}{\bmf}
\safemath{\vecg}{\bmg}
\safemath{\vech}{\bmh}
\safemath{\veci}{\bmi}
\safemath{\vecj}{\bmj}
\safemath{\veck}{\bmk}
\safemath{\vecl}{\bml}
\safemath{\vecm}{\bmm}
\safemath{\vecn}{\bmn}
\safemath{\veco}{\bmo}
\safemath{\vecp}{\bmp}
\safemath{\vecq}{\bmq}
\safemath{\vecr}{\bmr}
\safemath{\vecs}{\bms}
\safemath{\vect}{\bmt}
\safemath{\vecu}{\bmu}
\safemath{\vecv}{\bmv}
\safemath{\vecw}{\bmw}
\safemath{\vecx}{\bmx}
\safemath{\vecy}{\bmy}
\safemath{\vecz}{\bmz}

\safemath{\veczero}{\bmzero}
\safemath{\vecone}{\bmone}
\safemath{\vecxi}{\bmxi}
\safemath{\veclambda}{\bmlambda}
\safemath{\vecmu}{\bmmu}
\safemath{\vectheta}{\bmtheta}
\safemath{\vecphi}{\bmphi}
\safemath{\vecdelta}{\bmdelta}

\safemath{\matA}{\bA}
\safemath{\matB}{\bB}
\safemath{\matC}{\bC}
\safemath{\matD}{\bD}
\safemath{\matE}{\bE}
\safemath{\matF}{\bF}
\safemath{\matG}{\bG}
\safemath{\matH}{\bH}
\safemath{\matI}{\bI}
\safemath{\matJ}{\bJ}
\safemath{\matK}{\bK}
\safemath{\matL}{\bL}
\safemath{\matM}{\bM}
\safemath{\matN}{\bN}
\safemath{\matO}{\bO}
\safemath{\matP}{\bP}
\safemath{\matQ}{\bQ}
\safemath{\matR}{\bR}
\safemath{\matS}{\bS}
\safemath{\matT}{\bT}
\safemath{\matU}{\bU}
\safemath{\matV}{\bV}
\safemath{\matW}{\bW}
\safemath{\matX}{\bX}
\safemath{\matY}{\bY}
\safemath{\matZ}{\bZ}
\safemath{\matzero}{\bmzero}

\safemath{\matDelta}{\bDelta}
\safemath{\matLambda}{\bLambda}
\safemath{\matPhi}{\bPhi}
\safemath{\matSigma}{\bSigma}
\safemath{\matOmega}{\bOmega}
\safemath{\matTheta}{\bTheta}

\safemath{\matidentity}{\matI}
\safemath{\matone}{\matO}

\safemath{\rnda}{A}
\safemath{\rndb}{B}
\safemath{\rndc}{C}
\safemath{\rndd}{D}
\safemath{\rnde}{E}
\safemath{\rndf}{F}
\safemath{\rndg}{G}
\safemath{\rndh}{H}
\safemath{\rndi}{I}
\safemath{\rndj}{J}
\safemath{\rndk}{K}
\safemath{\rndl}{L}
\safemath{\rndm}{M}
\safemath{\rndn}{N}
\safemath{\rndo}{O}
\safemath{\rndp}{P}
\safemath{\rndq}{Q}
\safemath{\rndr}{R}
\safemath{\rnds}{S}
\safemath{\rndt}{T}
\safemath{\rndu}{U}
\safemath{\rndv}{V}
\safemath{\rndw}{W}
\safemath{\rndx}{X}
\safemath{\rndy}{Y}
\safemath{\rndz}{Z}

\safemath{\rveca}{\bimA}
\safemath{\rvecb}{\bimB}
\safemath{\rvecc}{\bimC}
\safemath{\rvecd}{\bimD}
\safemath{\rvece}{\bimE}
\safemath{\rvecf}{\bimF}
\safemath{\rvecg}{\bimG}
\safemath{\rvech}{\bimH}
\safemath{\rveci}{\bimI}
\safemath{\rvecj}{\bimJ}
\safemath{\rveck}{\bimK}
\safemath{\rvecl}{\bimL}
\safemath{\rvecm}{\bimM}
\safemath{\rvecn}{\bimN}
\safemath{\rveco}{\bomO}
\safemath{\rvecp}{\bimP}
\safemath{\rvecq}{\bimQ}
\safemath{\rvecr}{\bimR}
\safemath{\rvecs}{\bimS}
\safemath{\rvect}{\bimT}
\safemath{\rvecu}{\bimU}
\safemath{\rvecv}{\bimV}
\safemath{\rvecw}{\bimW}
\safemath{\rvecx}{\bimX}
\safemath{\rvecy}{\bimY}
\safemath{\rvecz}{\bimZ}

\safemath{\rvecxi}{\bmxi}
\safemath{\rveclambda}{\bmlambda}
\safemath{\rvecmu}{\bmmu}
\safemath{\rvectheta}{\bmtheta}
\safemath{\rvecphi}{\bmphi}

\safemath{\rmatA}{\bimA}
\safemath{\rmatB}{\bimB}
\safemath{\rmatC}{\bimC}
\safemath{\rmatD}{\bimD}
\safemath{\rmatE}{\bimE}
\safemath{\rmatF}{\bimF}
\safemath{\rmatG}{\bimG}
\safemath{\rmatH}{\bimH}
\safemath{\rmatI}{\bimI}
\safemath{\rmatJ}{\bimJ}
\safemath{\rmatK}{\bimK}
\safemath{\rmatL}{\bimL}
\safemath{\rmatM}{\bimM}
\safemath{\rmatN}{\bimN}
\safemath{\rmatO}{\bimO}
\safemath{\rmatP}{\bimP}
\safemath{\rmatQ}{\bimQ}
\safemath{\rmatR}{\bimR}
\safemath{\rmatS}{\bimS}
\safemath{\rmatT}{\bimT}
\safemath{\rmatU}{\bimU}
\safemath{\rmatV}{\bimV}
\safemath{\rmatW}{\bimW}
\safemath{\rmatX}{\bimX}
\safemath{\rmatY}{\bimY}
\safemath{\rmatZ}{\bimZ}

\safemath{\rmatDelta}{\bimDelta}
\safemath{\rmatLambda}{\bimLambda}
\safemath{\rmatPhi}{\bimPhi}
\safemath{\rmatSigma}{\bimSigma}
\safemath{\rmatOmega}{\bimOmega}
\safemath{\rmatTheta}{\bimTheta}

%% file: standard-macros.tex
\usepackage{amssymb}
\usepackage{amsfonts}
\usepackage{mathrsfs}
\usepackage{xspace}
\usepackage{bm}
\usepackage{fancyref}
\usepackage{textcomp}

\usepackage{multirow}
\usepackage{stmaryrd}

\newenvironment{textbmatrix}{	\setlength{\arraycolsep}{2.5pt}%
								\big[\begin{matrix}}{\end{matrix}\big]%
								\raisebox{0.08ex}{\vphantom{M}}}

\def\be{\begin{equation}}
\def\ee{\end{equation}}
\def\een{\nonumber \end{equation}}
\def\mat{\begin{bmatrix}}
\def\emat{\end{bmatrix}}
\def\btm{\begin{textbmatrix}}
\def\etm{\end{textbmatrix}}

\def\ba#1\ea{\begin{align}#1\end{align}}
\def\bas#1\eas{\begin{align*}#1\end{align*}}
\def\bs#1\es{\begin{split}#1\end{split}}
\def\bg#1\eg{\begin{gather}#1\end{gather}}
\def\bml#1\eml{\begin{multline}#1\end{multline}}
\def\bi#1\ei{\begin{itemize}#1\end{itemize}}

\newcommand{\lefto}{\mathopen{}\left}

\DeclareMathOperator*{\argmin}{arg\;min}		
\DeclareMathOperator{\Exop}{\opE}			

\newcommand{\Ex}[2]{\ensuremath{\Exop_{#1}\lefto[#2\right]}} 	

\safemath{\dirac}{\delta}					
\safemath{\krond}{\dirac}					

\safemath{\upto}{\uparrow}
\safemath{\downto}{\downarrow}
\safemath{\iu}{j}							
\safemath{\ev}{\lambda}						
\safemath{\hilseqspace}{l^{2}}				
\newcommand{\banachfunspace}[1]{\setL^{#1}}	
\safemath{\hilfunspace}{\banachfunspace{2}}	

\safemath{\SNR}{\textit{SNR}} 				
\safemath{\PAR}{\textit{PAR}} 				
\safemath{\No}{N_0}							
\safemath{\Es}{E_s}							
\safemath{\Eb}{E_b}							
\safemath{\EbNo}{\frac{\Eb}{\No}}
\safemath{\EsNo}{\frac{\Es}{\No}}

\DeclareMathOperator{\CHop}{\ensuremath{\opH}} 
\safemath{\tvir}{\rndh_{\CHop}}				
\safemath{\tvtf}{\rndl_{\CHop}}				
\safemath{\spf}{\rnds_{\CHop}}				
\safemath{\bff}{H_{\CHop}}					

\safemath{\ircf}{r_{h}}						
\safemath{\tftvcf}{r_{s}}					
\safemath{\tfcf}{r_{l}}						
\safemath{\bfcf}{r_{H}}						

\safemath{\tcorr}{c_h}						
\safemath{\scf}{c_{s}}						
\safemath{\tfcorr}{c_{l}}					
\safemath{\fcorr}{c_{H}}						

\safemath{\mi}{I}							
\safemath{\capacity}{C}						

\safemath{\normal}{\mathcal{N}}			
\safemath{\jpg}{\mathcal{CN}}			
\safemath{\mchain}{\leftrightarrow}		

\safemath{\dB}{\,\mathrm{dB}}
\safemath{\dBm}{\,\mathrm{dBm}}
\safemath{\Hz}{\,\mathrm{Hz}}
\safemath{\kHz}{\,\mathrm{kHz}}
\safemath{\MHz}{\,\mathrm{MHz}}
\safemath{\GHz}{\,\mathrm{GHz}}
\safemath{\s}{\,\mathrm{s}}
\safemath{\ms}{\,\mathrm{ms}}
\safemath{\mus}{\,\mathrm{\text{\textmu}s}}
\safemath{\ns}{\,\mathrm{ns}}
\safemath{\ps}{\,\mathrm{ps}}
\safemath{\meter}{\,\mathrm{m}}
\safemath{\mm}{\,\mathrm{mm}}
\safemath{\cm}{\,\mathrm{cm}}
\safemath{\m}{\,\mathrm{m}}
\safemath{\W}{\,\mathrm{W}}
\safemath{\mW}{\, \mathrm{mW}}
\safemath{\J}{\,\mathrm{J}}
\safemath{\K}{\,\mathrm{K}}
\safemath{\bit}{\,\mathrm{bit}}
\safemath{\nat}{\,\mathrm{nat}}

\safemath{\define}{\triangleq}			

\safemath{\equivalent}{\sim}
\safemath{\distas}{\sim}					
\safemath{\sdiff}{\Delta}				

\safemath{\reals}{\mathbb{R}}
\safemath{\positivereals}{\reals_{+}}
\safemath{\integers}{\mathbb{Z}}
\safemath{\posint}{\integers_{+}}
\safemath{\naturals}{\mathbb{N}}
\safemath{\posnaturals}{\naturals_{+}}
\safemath{\complexset}{\mathbb{C}}
\safemath{\rationals}{\mathbb{Q}}

\newcommand*{\fancyrefapplabelprefix}{app}		
\newcommand*{\fancyrefthmlabelprefix}{thm}		
\newcommand*{\fancyreflemlabelprefix}{lem}		
\newcommand*{\fancyrefcorlabelprefix}{cor}		
\newcommand*{\fancyrefdeflabelprefix}{def}		
\newcommand*{\fancyrefproplabelprefix}{prop}		
\newcommand*{\fancyrefexmpllabelprefix}{exmpl}
\newcommand*{\fancyrefalglabelprefix}{alg}		
\newcommand*{\fancyreftbllabelprefix}{tbl}		

\frefformat{vario}{\fancyrefseclabelprefix}{Section~#1}
\frefformat{vario}{\fancyrefthmlabelprefix}{Theorem~#1}
\frefformat{vario}{\fancyreftbllabelprefix}{Table~#1}
\frefformat{vario}{\fancyreflemlabelprefix}{Lemma~#1}
\frefformat{vario}{\fancyrefcorlabelprefix}{Corollary~#1}
\frefformat{vario}{\fancyrefdeflabelprefix}{Definition~#1}
\frefformat{vario}{\fancyreffiglabelprefix}{Figure~#1}
\frefformat{vario}{\fancyrefapplabelprefix}{Appendix~#1}
\frefformat{vario}{\fancyrefeqlabelprefix}{(#1)}
\frefformat{vario}{\fancyrefproplabelprefix}{Proposition~#1}
\frefformat{vario}{\fancyrefexmpllabelprefix}{Example~#1}
\frefformat{vario}{\fancyrefalglabelprefix}{Algorithm~#1}

%% file: defs.tex
 \newtheorem{thm}{Theorem}
 \newtheorem{cor}[thm]{Corollary}   

\safemath{\dictab}{[\,\dicta\,\,\dictb\,]}

\safemath{\ysig}{\bmy}
\safemath{\ysighat}{\hat{\ysig}}
\safemath{\ysigdim}{M}
\safemath{\xsig}{\bmx}
\safemath{\xsigdim}{N}
\safemath{\nx}{n_x}
\safemath{\zsig}{\bmz}
\safemath{\zsigdim}{\ysigdim}
\safemath{\rsig}{\bmr}
\safemath{\Adict}{\bA}
\safemath{\Adicttilde}{\widetilde{\Adict}}
\safemath{\Adictdim}{\outputdim\times\xsigdim}
\safemath{\avec}{\bma}
\safemath{\avectilde}{\tilde{\avec}}
\safemath{\Bdict}{\bB}
\safemath{\Bdicttilde}{\widetilde{\Bdict}}
\safemath{\Cdict}{\bC}
\safemath{\cvec}{\bmc}
\safemath{\Ddict}{\bD}
\safemath{\Ddictdim}{\ysigdim\times\xsigdim}
\safemath{\dvec}{\bmd}
\safemath{\Ddicttilde}{\widetilde{\bD}}
\safemath{\Bonb}{\bB}
\safemath{\bvec}{\bmb}
\safemath{\Bonbdim}{\ysigdim\times\ysigdim}
\safemath{\noise}{\bmn}
\safemath{\noisedim}{\ysigim}
\safemath{\err}{\bme}
\safemath{\errdim}{\ysigdim}
\safemath{\errset}{\setE}
\safemath{\nerr}{n_e}
\safemath{\delop}{\bP_\errset}
\safemath{\delopc}{\bP_{{\errset}^c}}

\safemath{\cplxi}{\imath}
\safemath{\cplxj}{\jmath}

\safemath{\dict}{\matD}
\safemath{\inputdim}{N}		
\safemath{\outputdim}{M}		
\safemath{\sparsity}{S}	
\safemath{\inputdimA}{{N_a}}	
\safemath{\inputdimB}{{N_b}}	
\safemath{\elemA}{{n_a}}	
\safemath{\elemB}{{n_b}}	
\safemath{\resA}{\matR_a}	
\safemath{\resB}{\matR_b}	
\safemath{\subD}{\matS} 
\safemath{\subA}{\matS_a} 
\safemath{\subB}{\matS_b} 
\safemath{\dicta}{\matA} 	
\safemath{\dictb}{\matB} 	
\safemath{\hollowS}{H}
\safemath{\hollowA}{H_a}
\safemath{\hollowB}{H_b}
\safemath{\cross}{Z}
\safemath{\coh}{\mu_d}			
\safemath{\coha}{\mu_a}			
\safemath{\cohb}{\mu_b}			
\safemath{\mubs}{\nu}	
\safemath{\cohm}{\mu_m} 
\safemath{\dictset}{\setD}	
\safemath{\dictsetp}{\dictset(\coh,\coha,\cohb)}	
\safemath{\dictsetgen}{\dictset_\text{gen}}
\safemath{\dictsetgenp}{\dictsetgen(\coh)}
\safemath{\dictsetonb}{\dictset_\text{onb}}
\safemath{\dictsetonbp}{\dictsetonb(\coh)}

\safemath{\leftside}{U}
\safemath{\rightsideA}{R_a}
\safemath{\rightsideB}{R_b}

\safemath{\indexS}{\setI_S} 

\safemath{\na}{n_a}			
\safemath{\nb}{n_b}			
\safemath{\coeffa}{p_i}	
\safemath{\coeffb}{q_j}	
\safemath{\seta}{\setP}		
\safemath{\setb}{\setQ}     
\safemath{\setw}{\setW}	
\safemath{\setz}{\setZ}	
\safemath{\cola}{\veca}		
\safemath{\colb}{\vecb}		
\safemath{\cold}{\vecd}		
\safemath{\inputvec}{\vecx} 	
\safemath{\error}{\vece}	
\safemath{\noiseout}{\vecz} 	
\safemath{\inputvecel}{x}
\safemath{\inputveca}{\vecx_a}
\safemath{\inputvecb}{\vecx_b}
\safemath{\outputvec}{\vecy}	
\safemath{\lambdamin}{\lambda_{\mathrm{min}}}

\safemath{\elltwo}{\ell_2}
\safemath{\ellone}{\ell_1}
\safemath{\ellzero}{\ell_0}
\safemath{\ellinf}{\ell_\infty}
\safemath{\ellinftilde}{\ell_{\widetilde\infty}}
\safemath{\licard}{Z(\coh,\coha,\cohb)}
\safemath{\xsol}{\hat{x}}
\safemath{\xbord}{x_b}		
\safemath{\xstat}{x_s}		
\safemath{\xstatLone}{\tilde{x}_s}
\safemath{\order}{\mathcal{O}} 
\safemath{\scales}{\Theta} 
\safemath{\ones}{\mathbf{1}} 
\safemath{\zeroes}{\mathbf{0}} 
\safemath{\thlone}{\kappa(\coh,\cohb)} 
\safemath{\constoneA}{\delta} 
\safemath{\constoneB}{\epsilon} 
\safemath{\nlarge}{L}				   
\safemath{\sumlarge}{S_\nlarge}
\safemath{\maxlarger}{P_\nlarge}	   
\safemath{\Pzero}{\textrm{P0}}	
\safemath{\Pone}{\textrm{P1}}
\safemath{\vecfir}{\vecw}			 
\safemath{\vecsec}{\vecz}
\safemath{\elvecfir}{w}              
\safemath{\elvecsec}{z}				 
\safemath{\nlargefir}{n}
\safemath{\normout}{\gamma}
\safemath{\auxfun}{h}
\safemath{\supp}{\textrm{supp}}

\safemath{\indexa}{\ell}
\safemath{\indexb}{r}
\safemath{\indexc}{i}
\safemath{\indexd}{j}

\safemath{\project}{P}

%% file: 0-abstract.tex
\begin{abstract}
Millimeter-wave (mmWave) communication in combination with massive multiuser multiple-input multiple-output (MU-MIMO) enables high-bandwidth data transmission to multiple users in the same time-frequency resource. 
The strong path loss of wave propagation at such high frequencies necessitates accurate channel state information to ensure reliable data transmission. 
We propose a novel channel estimation algorithm called BEAmspace CHannel EStimation (BEACHES), which leverages the fact that wave propagation at mmWave frequencies is predominantly directional. BEACHES adaptively denoises the channel vectors in the beamspace domain using \revision{an adaptive shrinkage procedure} that relies on Stein's unbiased risk estimator (SURE).
Simulation results for line-of-sight (LoS) and non-LoS mmWave channels reveal that BEACHES performs on par with state-of-the-art channel estimation methods while requiring orders-of-magnitude lower complexity.
To demonstrate the effectiveness of BEACHES in practice, we develop a very large-scale integration (VLSI) architecture and provide field-programmable gate array (FPGA) implementation results. 
Our results show that \revision{adaptive channel denoising} can be performed at high throughput and in a hardware-friendly manner for massive MU-MIMO mmWave systems with hundreds of antennas. 
\end{abstract}

\begin{IEEEkeywords}
Millimeter wave (mmWave), massive multiuser MIMO, channel estimation, nonparametric denoising, beamspace, Stein's unbiased risk estimator (SURE), very large-scale integration (VLSI), field-programmable gate array (FPGA).
\end{IEEEkeywords}


%% file: 1-introduction.tex
\section{Introduction}

Millimeter-wave (mmWave) communication \cite{rappaport13a, swindlehurst14a} and massive multiuser (MU) multiple-input multiple-output (MIMO)~\cite{rusek14a, larsson14a} are expected to be core technologies of next-generation wireless communication systems.
By combining both of these technologies, one can achieve unprecedentedly high-bandwidth data transmission to multiple user equipments (UEs) in the same time-frequency resource via fine-grained beamforming. 
The strong path loss of wave propagation at mmWave frequencies necessitates the infrastructure basestations (BSs) to acquire accurate channel state information (CSI) in order to perform data detection in the uplink (UEs transmit to BS) and MU precoding in the downlink (BS transmits to UEs) \cite{rappaport15b, gao16}.
To optimally determine the beamforming weights, accurate CSI is not only of paramount importance for hybrid analog-digital BS architectures~\cite{alkhateeb15a, sohrabi16, chen17} but also for emerging all-digital BS architectures~\cite{dutta2019case, panagiotis20}. 
In addition, the trend towards BS architectures with  low-precision data converters to reduce power consumption, interconnect bandwidth, and system costs~\cite{jacobsson17b, li17b, mo16b} requires novel algorithms and hardware designs that denoise the estimated channel vectors.

\subsection{Sparsity-Based Channel  Estimation}  \label{sec:lit_review}
Fortunately, wave propagation at mmWave frequencies is predominantly directional and real-world channels typically comprise only a small number of strong propagation paths, such as a line-of-sight (LoS) component and a few first-order reflections~\cite{rappaport15a}. These properties enable the design of sparsity-exploiting CSI estimation algorithms that effectively suppress channel estimation errors~\cite{alkhateeb14a,mo14b,schniter14a,deng18}.
Compressive sensing (CS)-based methods have been proposed for mmWave channel estimation in \cite{wang17, lee16}, including methods that rely upon orthogonal matching pursuit (OMP) \cite{lee16,MaechlerThesis, tsai18}.
The majority of such methods uses a discretization procedure of the number of propagation paths that can be resolved in the \emph{beamspace} (or angular) domain~\cite{brady13}, which results in a problem widely known as \emph{basis mismatch}~\cite{tang13}. 
To avoid the basis mismatch problem, sparse channel estimation for mmWave channels can, for example, be accomplished with atomic norm minimization (ANM)~\cite{bhaskar13, zhang15} or Newtonized OMP \cite{mamandipoor16}. ANM estimates a discrete set of propagation paths off-the-grid by solving a semidefinite program (SDP). Newtonized OMP (NOMP) is a more efficient alternative to ANM and iteratively refines the incident angles  of the dominant propagation paths off-the-grid with a complexity only slightly higher than that of conventional OMP. 
Although both of these methods do not suffer from the basis mismatch problem and exhibit excellent denoising performance,
they entail high computational complexity. Hence, from a hardware-implementation perspective, such methods are less attractive, especially in massive MU-MIMO systems where the complexity is dominated by the large number of BS antennas. 
In addition, the performance of both of these methods strongly depends on algorithm parameters that need to be tuned for the given propagation conditions. 

Another strain of sparsity-exploiting channel-estimation methods build upon approximate message passing (AMP) \cite{bellili19, huang19b}. While such methods promise high estimation accuracy, they suffer from a number of drawbacks when implemented in VLSI.
AMP-based methods require at least two matrix-vector multiplications in each iteration, whose dimension scales with the number of BS antennas, the number of UEs, and the pilot sequence length. In addition, each iteration requires multiple divisions  and other nonlinear  functions (such as exponentials and Q-functions).
As shown in \cite{jeon19b}, the presence of such nonlinear functions in AMP-based algorithms causes finite-precision issues when implemented with fixed-point arithmetic.

\revisionS{Sparsity has been exploited in many other applications in communication systems, including beam selection in mmWave systems~\cite{yeh18},  channel estimation for angle-division multiple access~\cite{liu19}, and sparse signal recovery via compressive sensing~\cite{somappa20}. Even though these results are not directly related to channel estimation in mmWave systems, the proposed adaptive denoising approach might find use in such applications.}

\subsection{Contributions}
In order to perform denoising-based channel estimation in real-world systems,  we propose a low-complexity and \revision{adaptive channel estimation} algorithm for massive MU-MIMO mmWave systems that can be implemented efficiently in VLSI. 
Our main contributions are summarized as follows:
\begin{itemize}
\item We propose a novel channel estimation algorithm that relies on Stein's unbiased risk estimator (SURE), which we call BEAmspace CHannel EStimation (BEACHES). BEACHES exploits sparsity of mmWave channels in the beamspace domain and adaptively denoises the channel vectors at a fixed computational complexity that scales with  $O(B\log(B))$, where~$B$ is the number of BS antennas.
\item We prove that BEACHES minimizes the mean-square error (MSE) between the noiseless and denoised channel vector in the large-antenna limit, i.e., when  $B\to\infty$, without requiring tedious parameter tuning. 
\item We evaluate the efficacy of BEACHES for LoS and non-LoS mmWave channel models and show that it performs on par with state-of-the-art channel estimation algorithms in terms of uncoded bit error-rate, but at orders-of-magnitude lower computational complexity. 
\item We develop a very large-scale integration (VLSI) architecture and present corresponding field-programmable gate array (FPGA) implementation results, which demonstrate that BEACHES enables high-throughput channel estimation in a hardware-efficient manner. 
\end{itemize}

\subsection{Notation}
Lowercase and uppercase boldface letters designate column vectors and matrices, respectively. 
For a vector $\bma$, the $k$th entry is denoted by~$[\bma]_k=a_k$; the real and imaginary parts are indicated with $[\bma]_{\mathcal{R}}=\bma_{\mathcal{R}}$ and $[\bma]_{\mathcal{I}}=\bma_{\mathcal{I}}$, respectively. The $\ell_1$-norm and $\ell_2$-norm of a vector $\bma$ is~$\|\veca\|_1$ and $\|\veca\|_2$, respectively. 
For a matrix $\bA$, we define its transpose and conjugate transpose as $\bA^\Tran$ and $\bA^\Herm$, respectively. 
The $N\times M$ all-zeros,  $N\times N$ identity, and $N\times N$ discrete Fourier transform (DFT) matrices are $\bZero_{N\times M}$, $\bI_N$, and $\bF$, respectively; the DFT matrix is normalized so that $\bF\bF^H=\bI_N$.
Vectors in the DFT domain are designated with a hat as in $\hat{\bma}=\bF\bma$. 
A proper complex-valued Gaussian vector  $\bma$ with mean vector~$\bmm$ and covariance matrix~$\bK$ is written as \mbox{$\bma \sim \setC\setN(\bmm,\bK)$} and its probability density function (PDF) as $f^{\setC\setN}(\bma;\bmm,\bK)$. 
A real-valued Gaussian vector  $\bma$ with mean vector~$\bmm$ and covariance matrix~$\bK$ is written as \mbox{$\bma \sim \setN(\bmm,\bK)$} and its PDF as $f^{\setN}(\bma;\bmm,\bK)$. 
The expectation operator is $\Ex{}{\cdot}$.  
Optimal values are designated with the superscript $^\star$.

\subsection{Paper Outline}
The rest of the paper is organized as follows. \fref{sec:systemmodel} introduces the system model and outlines the concept of denoising-based beamspace channel estimation. \fref{sec:beaches} details the BEACHES algorithm and presents the simulation results. \fref{sec:implementation} proposes a VLSI architecture and provides FPGA implementation results. We conclude in \fref{sec:conclusions}. All proofs are relegated to the appendices.
%

%% file: 2-system_model.tex
\section{System Model}
\label{sec:systemmodel}

We now introduce the system model and summarize existing methods that perform beamspace channel estimation.
\begin{figure}[tp]
\centering
\includegraphics[width=0.95\columnwidth]{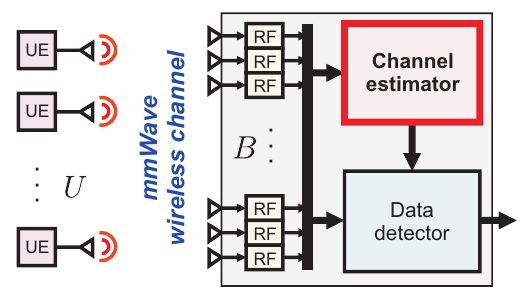}
\caption{Considered massive MU-MIMO mmWave uplink system: $U$ user equipments (UEs) transmit pilots over a mmWave wireless channel, which are used to estimate the channel vectors associated to each UE at the $B$-antenna basestation. This paper focuses on computationally efficient methods that denoise the measured channel vectors in the channel estimator unit. }
\label{fig:system_overview}
\end{figure}
\subsection{System Model}
We consider a massive MU-MIMO mmWave uplink system as illustrated in \fref{fig:system_overview}. The BS is equipped with $B$ antennas arranged as a uniform linear array (ULA) and communicates with~$U$ single-antenna UEs in the same time frequency resource.\footnote{An extension of our algorithm and hardware designs to two-dimensional BS antenna arrays is part of ongoing work.} 
We focus on pilot-based channel estimation, i.e., where the UEs transmit orthogonal pilots in a dedicated training phase and the BS estimates the propagation paths between the UEs and the BS antenna array. 
Assuming flat-fading channel conditions, the BS estimates the $B$-dimensional complex channel vector $\bmh\in\complexset^B$ for each UE. 
Furthermore, by assuming that (i) wave propagation is predominantly directional, which is valid if the wavelength is much smaller than the objects interacting with the waves~\cite{akdeniz14a,rappaport15b}, and (ii) the distance between UE (as well as the scatterers)  and BS is sufficiently large, we can use the following well-known plane-wave approximation to model wave propagation at mmWave frequencies from a given UE to the BS \cite{tse05a}:
 \begin{align} \label{eq:planewavemodel}
\bmh = \sum_{\ell=1}^{L} \alpha_\ell \bma(\Omega_\ell), \,\, \bma(\Omega) = \big[e^{j0\Omega},e^{j1\Omega},\ldots,e^{j(B-1)\Omega} \big]^\Tran\!.
\end{align}
Here,  $L$ refers to the total number of paths arriving at the antenna array (including a potential line-of-sight path),  \mbox{$\alpha_\ell\in\complexset$} is the complex-valued channel gain of the $\ell$th path, and $\bma(\Omega_\ell)$ represents a complex-valued sinusoid containing the relative phases between BS antennas, where $\Omega_\ell\in[0,2\pi)$ is determined by the incident angle of the $\ell$th path to the antenna array. 
 
With pilot-based channel estimation methods, we only have access to noisy measurements of the channel vector $\bmh$. We model such noisy measurements in the antenna domain as 
\begin{align}\label{eq:antenna_model}
\bmy=\bmh+\bme,
\end{align} 
where $\bme \sim \setC\setN(\bZero_{B\times1},\Eo \bI_B)$ represents channel estimation error with variance $\Eo$ per complex entry. Note that for pilot-based channel estimation methods, the channel estimation errors are Gaussian and there is a linear relationship between $\Eo$ and the thermal noise variance $\No$; see \fref{sec:SimRes} for the details.

\begin{remark}
The channel model in \fref{eq:planewavemodel} is appropriate for flat-fading channels assuming UEs with a single transmit antenna. For UEs that are equipped with an antenna array but transmit a single stream (layer) via beamforming, the channel vectors can still be modeled as in \fref{eq:planewavemodel}. For channels that exhibit frequency selectivity, we can consider orthogonal frequency-division multiplexing (OFDM), where each subcarrier is associated with a channel vector as in \fref{eq:planewavemodel}. For single-carrier (SC) transmission in frequency-selective channels or UEs that transmit multiple streams concurrently, multiple channel vectors would need to be estimated (one for each tap in the impulse response and for each layer). An analysis of this scenario is ongoing work. 
Finally, we emphasize that BEACHES continues to work if the channel vectors follow a more realistic propagation model than the one in \fref{eq:planewavemodel}. The simulation results provided in \fref{sec:SimRes} with mmWave channel models confirm this claim. 
\end{remark}

\begin{remark}
\revision{In what follows, we ignore system and hardware impairments, such as timing, frequency, and sampling rate offsets, I/Q imbalance, and analog-to-digital converter (ADC) nonlinearities. In cases where the aggregate effect of the residual hardware impairments can be modeled as Gaussian noise \cite{studer10b}, the model in \fref{eq:planewavemodel} remains valid.  For basestation architectures with 1-bit ADCs, a specialized version of BEACHES has been proposed recently in \cite{gallyas20}. The design of robust channel estimation algorithms for more specific system and hardware impairments is left for future work. }
\end{remark}

\subsection{Beamspace Representation}
\begin{figure}[tp]
\subfigure[LoS channel]{\includegraphics[width=0.495\columnwidth]{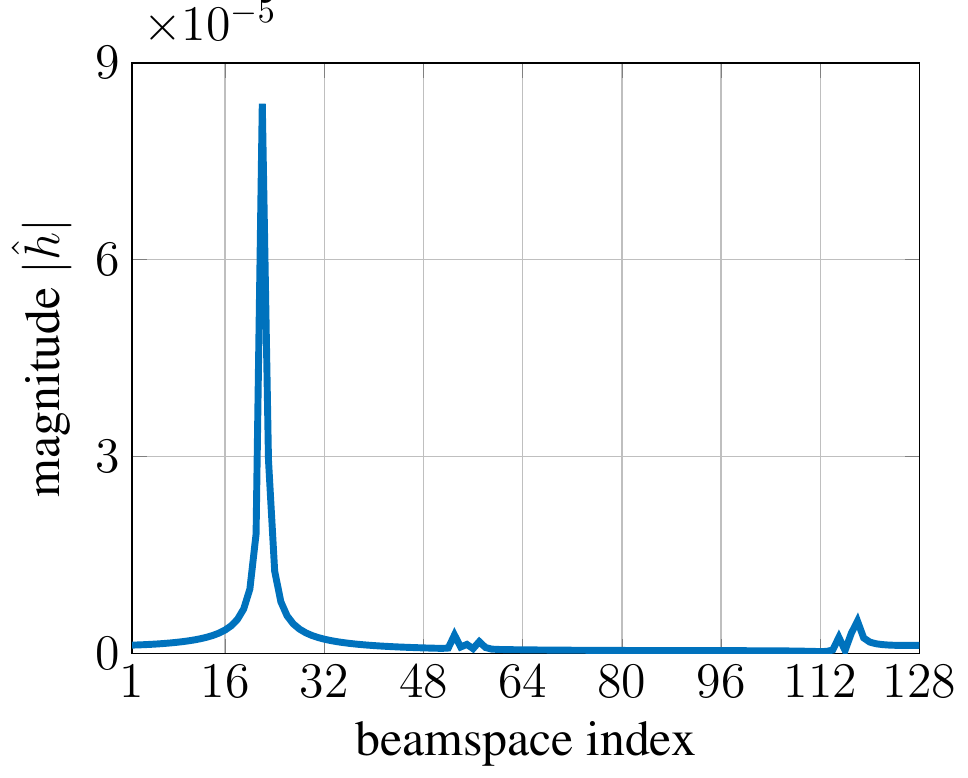}}
\subfigure[Non-LoS channel]{\includegraphics[width=0.495\columnwidth]{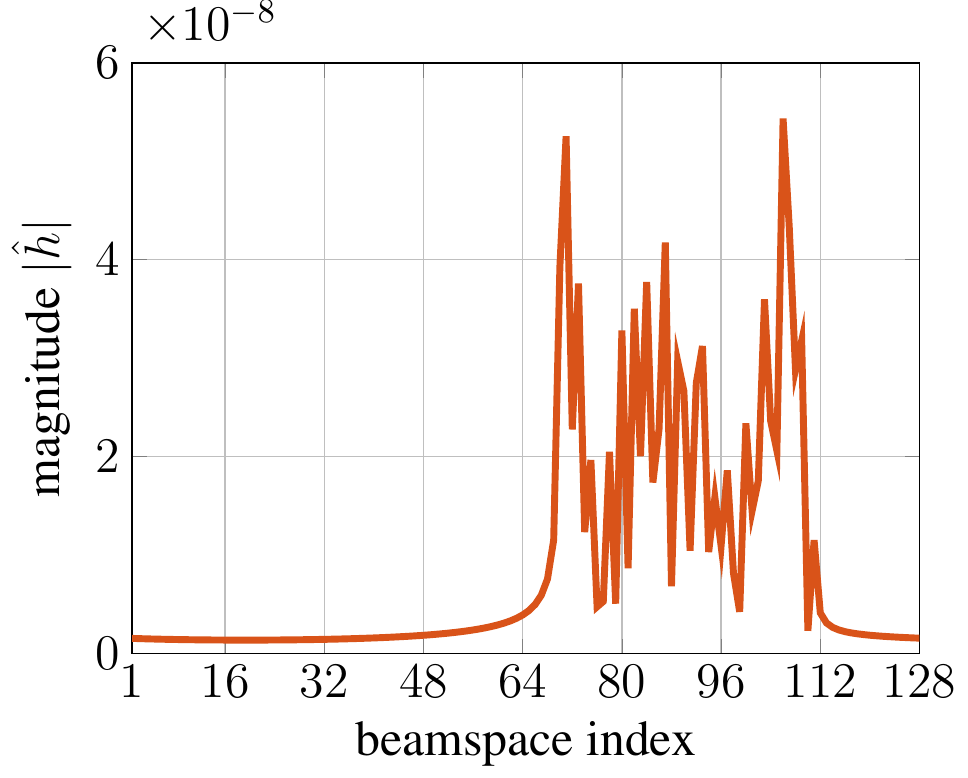}}
\caption{Examples of a line-of-sight (LoS) channel vector (a) and a non-LoS  channel vector (b) in the discrete beamspace domain. The channel vectors are generated with the mmMAGIC UMi model  at 60\,GHz for a $128$ antenna BS with a uniform-linear array (ULA) using $\lambda/2$ antenna spacing. One can clearly see the sparse nature of channel vectors in the beamspace domain.} 
\label{fig:sparse_channel}
\end{figure}

The  model in \fref{eq:planewavemodel} describes the channel vector in the \emph{antenna domain}, i.e., each entry of the channel vector $\bmh$ is associated with an antenna element in the BS array. 
Since the channel vectors~$\bmh$ are modeled as a superposition of $L$ complex-valued sinusoids, it is advantageous to transform the observed vector~$\bmy$ into the discrete Fourier transform domain according to  $\hat\bmy=\bF\bmy$, where~$\bF$ is the $B\times B$ DFT matrix.
This transformation is known to convert the noisy channel vector~$\bmy=\bmh+\bme$ into the so-called \emph{discrete beamspace domain} (also known as angular domain)~$\hat\bmy$, in which each entry is associated with a specific incident angle (with respect to the BS antenna array) \cite{brady13}. 
More importantly, if the number of paths~$L$ is significantly smaller than the number of BS antennas~$B$, then the beamspace representation~$\hat\bmh$ of the noiseless channel vector~$\bmh$ will be (approximately) sparse~\cite{schniter14a}. 
In other words, most of the channel vector's energy is concentrated on a few entries, which are associated with the indices corresponding to the angles of the arriving waves. 
This key property of the beamspace representation is illustrated in \fref{fig:sparse_channel}, which shows the magnitude of $\hat\bmh$ for noiseless LoS and non-LoS channel vectors generated with the QuaDRiGa mmMAGIC urban micro (UMi) model at a carrier frequency of 60\,GHz~\cite{QuaDRiGa_tech_rpt}.
For the LoS case in \fref{fig:sparse_channel}(a), we see that the channel vector consists of one strong LoS component and two weak first-order reflections arriving at two distinct angles.\footnote{Note that the strong signal arrived off-the-grid, which causes it  to be spread across multiple angular bins. This is an instance of the off-the-grid problem that has been studied extensively in the compressive sensing literature \cite{tang13}.}
For the non-LoS case in \fref{fig:sparse_channel}(b), we see that the arriving waves are (i) weaker than for the LoS case and (ii) spread across a wider range of angles. Nevertheless, the channel vector remains to be sparse in the non-LoS case.

\subsection{Channel Vector Denoising in the Beamspace Domain}

The sparse nature of mmWave channel vectors in the beamspace domain enables the use of  algorithms that denoise the channel vectors at the BS. 
The main idea behind such channel estimation methods is to first transform the observed noisy channel vector~$\bmy$ in the antenna domain~\fref{eq:antenna_model} to the beamspace domain according to
\begin{align} \label{eq:beamspace_model}
\hat{\bmy}=\bF\bmy =  \hat{\bmh}+\hat{\bme},
\end{align}
where $\hat{\bme}=\bF\bme$ has the same statistics as the antenna domain channel estimation error vector $\bme$.
It is then possible to exploit the fact that most of the arriving signal energy is concentrated on a few incident angles and to suppress noise associated with angles that do not pertain to the incoming signals.

To perform denoising, a variety of algorithms have been proposed in the literature (see also the discussion in \fref{sec:lit_review}). While most existing methods, such as OMP, suffer from the off-the-grid problem~\cite{tang13}, more sophisticated methods such as ANM~\cite{bhaskar13, zhang15} and NOMP~\cite{mamandipoor16}, avoid this problem by identifying the dominant paths in the continuous beamspace domain.
Unfortunately, such methods exhibit high computational complexity, especially for a large number of BS antennas~$B$, which prevents their use in real-time applications. 
We next introduce a nonparametric beamspace denoising algorithm that is computationally efficient, can be implemented in hardware, and performs on par with sophisticated off-the-grid beamspace channel estimation algorithms.  
 

%% file: 3-beaches.tex
\section{BEACHES: BEAmspace CHannel EStimation}
\label{sec:beaches}

We now introduce BEACHES, an efficient algorithm for channel vector denoising in the beamspace domain.

\subsection{Channel Vector Denoising via Soft-Thresholding}
\begin{figure}[tp]
	\centering
	\includegraphics[width=0.65\columnwidth]{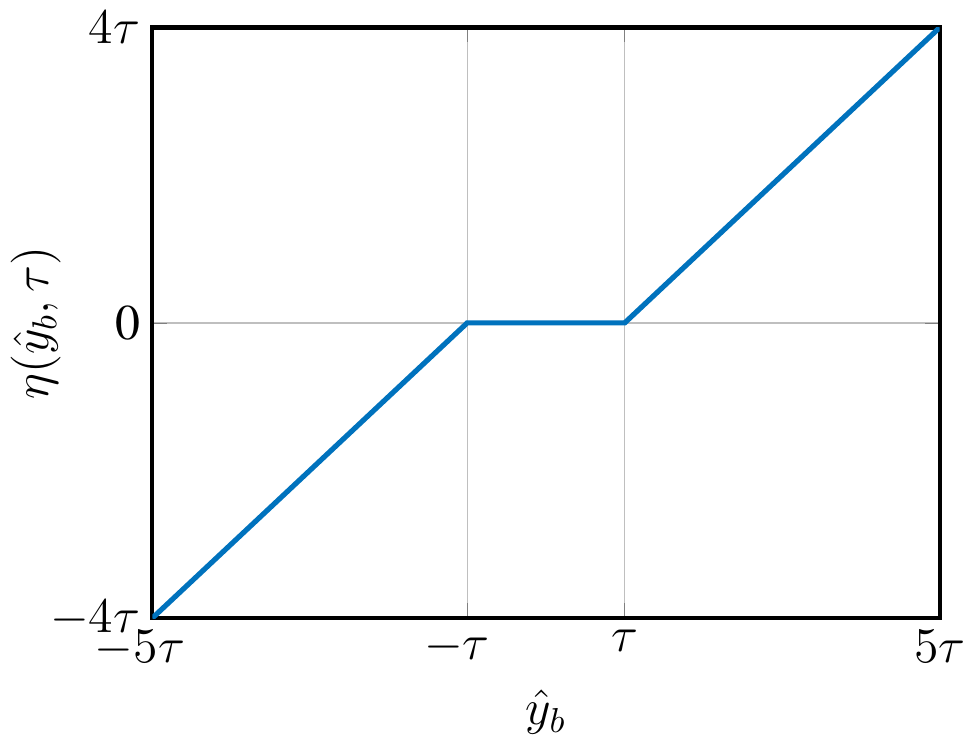}
	\vspace{-0.2cm}
	\caption{The soft-thresholding function ${\eta}(\yhate{b}, \tau)$ for real values $\yhate{b}$.}
	\label{fig:eta}
\end{figure}

The denoising and sparse signal recovery  literature~\cite{alkhateeb13,alkhateeb14a,mo14b,schniter14a,deng18} describes a number of algorithms that are suitable for channel-vector denoising in the beamspace domain.
The least absolute shrinkage and selection operator (LASSO) \cite{donoho95,tibshirani96,tropp10} is among the most popular  methods, which, in our application, corresponds to the following optimization problem: 
\begin{align} \label{eq:LASSO}
\eta(\hat\bmy,\tau) =\argmin_{\hhat^{\prime}\in\complexset^B} \frac{1}{2} \|\yhat-\hhat^{\prime}\|_2^2+\tau \|\hhat^{\prime}\|_1.
\end{align}
Here, we apply LASSO directly to the beamspace representation of the observed channel vector~\fref{eq:beamspace_model} and $\tau\in\reals_{+}$ is a carefully-chosen denoising parameter. 
A closed-form expression for the solution to \fref{eq:LASSO} in the complex case has been derived in~\cite[App.~A]{maleki13} and is given by the well-known \emph{soft-thresholding operator} $\eta(\yhat,\tau)$ defined entry-wise as 
\begin{align} \label{eq:softthresholdingfunction}
[\eta(\yhat,\tau)]_b=\frac{\yhate{b}}{|\yhate{b}|}\max{\{|\yhate{b}|-\tau,0\}}, \quad b=1,\ldots,B,
\end{align}
where we define $y/|y|=0$ for $y=0$. \fref{fig:eta} depicts the soft-thresholding function $\eta(\yhat,\tau)$, which  simply shrinks the magnitude of its input by $\tau$ 
{or} sets it to zero if the magnitude was smaller than $\tau$.

While soft-thresholding is widely used for denoising sparse signals, its performance strongly depends on the choice of the denoising parameter $\tau$ \cite{donoho95,mousavi13}.
Since the propagation conditions, such as the number of arriving paths (sparsity), the incident angles (locations of the nonzero components), and the received signal strength (magnitudes), can vary widely in wireless communication systems, the design of robust methods \revision{that} adaptively select the optimal denoising parameter is critical.
\revision{We now develop an adaptive approach that optimally tunes the denoising parameter~$\tau$ in a computationally-efficient manner.}

\begin{remark}
\revision{BEACHES only requires knowledge of the noise variance $\No$, which is typically known as it is determined by thermal noise originating in the receiver's RF circuitry.}
\end{remark}

\subsection{Computing the Optimal Denoising Parameter}
We are interested in computing the optimal denoising parameter $\tau^\star$ that minimizes the mean square error (MSE) between the denoised beamspace channel vector  
and the noiseless beamspace channel vector $\hat\bmh$, defined as follows:
\begin{align} \label{eq:MSE}
\textit{MSE} =  \frac{1}{B}\Ex{}{\|\eta(\yhat,\tau)-\hhat\|_2^2}\!.
\end{align}
In~\fref{eq:MSE}, expectation is with respect to $\hat\bmy$. 
In what follows, we denote the optimal denoised channel vector by 
$\hat\bmh^\star=\eta(\yhat,\tau^\star)$.

Unfortunately, determining the optimal denoising parameter~$\tau^\star$ that minimizes the MSE in \fref{eq:MSE} requires knowledge of the noiseless beamspace channel vector $\hhat$, which is unknown in practice.  
{To resolve this issue, we propose to minimize Stein's unbiased risk estimate (SURE) as a surrogate for the MSE.}

The following result provides an expression for SURE in the complex domain {and shows that SURE is an unbiased estimator of the MSE that is \emph{independent} of $\hhat$}. The proof of the following result is given in \fref{app:MSEtoSURE}.
\begin{thm} 
\label{thm:MSEappox}
Let $\hat\bmh \in  \complexset^{B}$ be an unknown vector and \mbox{$\yhat \in \complexset^{B}$} a noisy observation vector distributed as \mbox{$\hat\bmy\sim\setC\setN(\hat\bmh,\Eo\bI_B)$}.
Let $\mu(\hat{\bmy})$ be an estimator of~$\hat\bmh$ from $\yhat$ that is weakly differentiable and operates element-wise on vectors. Then, Stein's unbiased risk estimate given by
%
\begin{align} \label{eq:complexSURE}
\textit{SURE}=&\, \frac{1}{B}\|\mu(\yhat)-\yhat\|_2^2+\Eo\nonumber\\ 
&\,+\frac{\Eo}{B} \sum_{b=1}^{B} \left( \frac{\partial {[\mu_{\mathcal{R}}(\yhat)]_b}}{\partial [\yhat_{\mathcal{R}}]_b}+\frac{\partial {[\mu_{\mathcal{I}}(\yhat)]_b}}{\partial [\yhat_{\mathcal{I}}]_b}-2\right)\!,
\end{align}
is an unbiased estimate of the MSE, i.e., satisfies
\begin{align} \label{eq:MSEvsSURE}
\Ex{}{\textit{SURE}} = \textit{MSE}.
\end{align}
\end{thm}

By setting $\mu(\yhat)=\eta(\yhat,\tau)$, we can use \fref{thm:MSEappox} to obtain the following SURE expression for the soft-thresholding function. The  proof is given in \fref{app:shrinkageSURE}.
\begin{cor} \label{cor:surecoolandcrazy}	
	For the complex-valued soft-thresholding function $\mu(\yhat) = \eta(\yhat,\tau)$ in \fref{eq:softthresholdingfunction}, SURE in \fref{eq:complexSURE} is given by\footnote{As discussed in \fref{app:shrinkageSURE}, the value of $\text{SURE}_\tau$ is undefined for $\tau=|\yhate{b}|$, $b=1,\dots, B$, due to the non-differentiability of the function $\eta(\yhat,\tau)$.}
	\begin{align} 
	\nonumber
	\textit{SURE}_\tau = &\,
	\frac{1}{B}\sum_{b:|\yhate{b}|<\tau}|\yhate{b}|^2 + \frac{1}{B}\sum_{b:|\yhate{b}|>\tau}\tau^2 + \Eo\\\label{eq:shrinkageSURE}
	&\, - \frac{\Eo}{B}\tau\sum_{b:|\yhate{b}|>\tau} \frac{1}{|\yhate{b}|} - \frac{2\Eo}{B}\sum_{b:|\yhate{b}|<\tau}1.
	\end{align}
\end{cor}
The next result shows that the value of SURE in \fref{eq:shrinkageSURE} converges to the MSE given by \fref{eq:MSE} in the large antenna limit, i.e., for $B\to\infty$. The proof is given in \fref{app:SUREconvergence}.
\begin{thm} \label{thm:SUREconvergence}
In the large-antenna limit, where $B\to\infty$, $\textit{SURE}_\tau$ in \fref{eq:shrinkageSURE} converges to the MSE in \fref{eq:MSE}, i.e., we have
\begin{align} \label{eq:SUREconvergence}
\lim_{B\to\infty}\textit{SURE}_\tau = \textit{MSE}. 
\end{align}
\end{thm}
From  Theorems \ref{thm:MSEappox} and~\ref{thm:SUREconvergence} it is evident that SURE will be an accurate proxy for the MSE in massive MU-MIMO mmWave systems as $B$ is expected to be large. 
It is crucial to realize that the SURE expression in \fref{eq:shrinkageSURE} is \emph{independent} of the true beamspace channel vector $\hhat$. In fact, the result \fref{eq:shrinkageSURE} only depends on the magnitudes of the \emph{observed} beamspace channel vector $\yhat$, the channel estimation error variance~$\Eo$ (which is determined by the thermal noise variance $\No$), the number of BS antennas~$B$, and the denoising parameter~$\tau$. 
This insight combined with the two key properties in \fref{eq:MSEvsSURE} and \fref{eq:SUREconvergence} enables us to perform asymptotically-optimal MSE-based denoising by solving the SURE-based quantity 
\begin{align} \label{eq:optimalSUREvalue}
\tau^\star = \argmin_{\tau\in\reals_{+}}\,\textit{SURE}_\tau.
\end{align}
Unfortunately, no closed-form solution to this optimization problem is known. 
Reference~\cite{mousavi13} uses a bisection procedure to approximate the optimal value of a similar SURE expression in a sparse signal recovery  application. 
In stark contrast to such approximate methods, we next propose BEACHES, a hardware-friendly algorithm that computes the \emph{optimal} denoising parameter~$\tau^\star$ in \fref{eq:optimalSUREvalue} using a deterministic procedure whose complexity scales only with $O(B\log(B))$.

\begin{remark}
SURE-based denoising was put forward in \cite{donoho95} for wavelet denoising of real-valued signals. 
In \cite{upadhya14}, SURE has been applied for denoising complex-valued channel observation in OFDM-based single-antenna systems, exploiting sparsity of the impulse responses. \revision{The method in \cite{upadhya14} uses SURE} to find the coefficients of a frequency-domain filter, while the value of the  shrinkage threshold was determined empirically.
In contrast to these results, BEACHES  exploits sparsity in the beamspace domain and determines the optimal denoising parameter~$\tau^\star$ in $O(B\log(B))$ time. 
We note that BEACHES could  be combined with the method in \cite{upadhya14} in order to improve channel estimation in OFDM-based massive MU-MIMO mmWave systems. 
\end{remark}

%
\input{3c-algorithm}

\input{3d-simulation}

%% file: 3c-algorithm.tex

\subsection{The BEACHES Algorithm} \label{sec:BEACHES_alg}
Reference  \cite{donoho95} outlines an efficient procedure to minimize SURE for wavelet-denoising of real-valued signals. 
In what follows, we propose a similar strategy to minimize \fref{eq:shrinkageSURE} for the complex-valued case.
Instead of continuously sweeping the denoising parameter~$\tau$ through the interval $[0,\infty)$, we first sort the absolute values of the vector $\hat\bmy$ in ascending order and call the resulting sorted vector $\hat{\bmy}^s$. 
We then search for the optimal denoising parameter~$\tau$ only between each pair of consecutive elements of the sorted vector, i.e., $\tau\in\left(\hat{y}^s_{k-1}, \hat{y}^s_{k}\right)$ for $k=1,\ldots,B+1$, where we define $\hat{y}^s_{0}=0$ and $\hat{y}^s_{B+1}=\infty$ to account for the first interval $\left(0, \hat{y}^s_{1}\right)$, and last the interval $\left(\hat{y}^s_{B}, \infty \right)$.
In the $k$th interval, SURE in \fref{eq:shrinkageSURE} is a quadratic function of $\tau$ given by 
\begin{align}
\label{eq:sortedSURE}
\textit{SURE}_{\tau,k}  = &\,
	\frac{1}{B}\underbrace{\sum_{b=1}^{k-1}(\yshate{b})^2}_{=S} + \frac{(B-k+1)}{B}\tau^2 + \Eo \notag\\
	 &\,-\frac{\Eo}{B}\tau \underbrace{\sum_{b=k}^{B}(\yshate{b})^{-1}}_{=V} - \frac{2\Eo}{B}(k-1).
\end{align}
{For each index $k\in\{1,\ldots,B+1\}$, we compute the value of $\tau = \tau_k^\star$ that locally minimizes {$\text{SURE}_{\tau,k}$} in the interval $\tau\in(\yshate{k-1},\yshate{k})$.
Since SURE in \fref{eq:sortedSURE} is a quadratic function of~$\tau$, the minimal value in each interval is either at the minimum of the quadratic function \fref{eq:sortedSURE} or at one of the two interval boundaries\footnote{Note that $\text{SURE}_\tau$ and $\text{SURE}_{\tau,k}$ are not defined for $\tau=\yshate{k-1}$ and $\tau=\yshate{k}$. We evaluate $\text{SURE}_{\tau,k}$ for two values arbitrarily close to these boundaries, i.e., $\tau=\yshate{k-1}+\epsilon$ and $\tau=\yshate{k}-\epsilon$ where $\epsilon>0$ is small compared to $\tau$.}, i.e., $\yshate{k-1}$ or $\yshate{k}$. 
The minimum value of the expression in~\fref{eq:sortedSURE} is attained by
$\tau_k^Q = \frac{E_0}{2(B-k+1)} {\sum_{b=k}^{B}(\hat{y}^s_{b})^{-1}}$.
Since the function $\text{SURE}_{\tau,k}$ is convex within each interval $\left(\hat{y}^s_{k-1}, \hat{y}^s_{k}\right)$, the optimal parameter $\tau^\star_k$ in each interval $k=1,\ldots,B+1$, is given by
\begin{align} \label{eq:tauopt}
\tau^\star_k = \left\{\begin{array}{ll}
\, \tau_k^Q,  & \hat{y}^s_{k-1}<\tau_k^Q<\hat{y}^s_{k}, \\ 
\, \hat{y}^s_{k-1}, & \tau_k^Q<\hat{y}^s_{k-1}, \\ 
\, \hat{y}^s_{k}, & \tau_k^Q>\hat{y}^s_{k}, 
\end{array}\right.
\end{align}
or simply 
$\tau^\star_k = \max \{ \hat{y}^s_{k-1},  \min \{ \hat{y}^s_{k}, \tau_k^Q \} \}$.
After identifying the optimal value $\tau^\star_k$ in each interval, the parameter $\tau^\star$ that achieves the global minimum can be found 
by comparing all the local minima, i.e., by solving
\begin{align} \label{eq:tau_global_min}
\tau^\star=\argmin_{\tau^\star_k\,  , \, k=1,\ldots,B+1} \, {\textit{SURE}_{\tau_k^\star,k}}.
\end{align} 

Our procedure does not need to recalculate SURE in~\fref{eq:sortedSURE} from scratch while searching through $k=1,\ldots,B+1$. Instead, for each value of $k$, we sequentially  update the two quantities $S=\sum_{b=1}^{k-1}{(\hat{y}^s_{b})^2}$ and $V=\sum_{b=k}^{B}{(\hat{y}^s_{b})^{-1}}$, noting that the magnitudes of the vector~$\hat{\bmy}^s$ are sorted.
\fref{alg:BEACHES}, which we call BEACHES, exploits exactly this observation.
{Lines \ref{alg:BEACHES:line5} to~\ref{alg:BEACHES:line14} detail the search procedure {described in \fref{eq:tau_global_min}}; this part of the algorithm 
only involves scalar operations (additions, multiplications, divisions, and comparisons) all of which scale with ${O}(1)$.}
As a consequence, this iterative search has a complexity of only ${O}(B)$. If we assume that the DFT and inverse DFT in line \ref{alg:BEACHES:line3} and line~\ref{alg:BEACHES:line16} are carried out with a fast Fourier transform (FFT) and inverse FFT (IFFT), respectively, and the sorting procedure in line \ref{alg:BEACHES:line4} uses a fast sorting algorithm (e.g., merge sort) with complexity ${O}(B\log(B))$, then the complexity of BEACHES is ${O}(B\log(B))$. Furthermore, we emphasize that  sorting, FFT, iterative scan, and IFFT are all hardware friendly operations; see \fref{sec:implementation} for a corresponding VLSI design. A detailed complexity comparison of BEACHES to NOMP and ANM is provided in \fref{sec:complexityanalysis}.

\begin{algorithm}[t]
\caption{BEACHES: BEAmspace CHannel EStimation \label{alg:BEACHES}}
\begin{algorithmic}[1]
\STATE {\bf input} $\bmy$ and $\Eo$
\STATE $\hat\bmy = \text{FFT}(\bmy)$ \label{alg:BEACHES:line3}
\STATE $\yhat^s=\text{sort}\{|\yhat|,\text{`ascend'}\}$, {$\hat{y}^s_{0}=0$, and $\hat{y}^s_{B+1}=\infty$}  	\label{alg:BEACHES:line4}
\STATE  $S=0$, {$V=\sum_{k=1}^{B} {(|\yhate{k}|)^{-1}}$} and $\textit{SURE}_{\text{min}}=\infty$											\label{alg:BEACHES:line2}
\FOR{$k=1,\ldots,B+1$}															\label{alg:BEACHES:line5}
\STATE {$\tau_k^\star = \max \{ \hat{y}^s_{k-1},  \min \{ \hat{y}^s_{k}, \frac{E_0}{2(B-k+1)} V \} \}$} \label{alg:BEACHES:line6}
\STATE $\textit{SURE}_{\tau_k^\star,k} = \frac{S}{B} + \frac{(B-k+1)}{B}{\tau_k^\star}^2 + \Eo$ \label{alg:BEACHES:line7}

$\qquad \qquad \quad \,\, - \frac{\Eo}{B} \tau_k^\star  V - \frac{2\Eo}{B} (k-1)$

\IF{$\textit{SURE}_{\tau_k^\star,k}<\textit{SURE}_{\text{min}}$}													\label{alg:BEACHES:line8}
\STATE $\textit{SURE}_{\text{min}} = \textit{SURE}_{\tau_k^\star,k}$
\STATE $\tau^\star = \tau_k^\star$
\ENDIF																				\label{alg:BEACHES:line11}
\STATE $S = S + (\hat{y}^s_k)^2$
\STATE $V = V - (\hat{y}^s_k)^{-1}$															\label{alg:BEACHES:line13}
\ENDFOR																				\label{alg:BEACHES:line14}
\STATE ${\hat{h}^\star_k}=\frac{\hat{y}_k}{|\hat{y}_k|}\max{\{|\hat{y}_k|-\tau^\star,0\}}$, $k=1,\ldots,B$
\STATE $\bmh^\star = \text{IFFT}(\hat{\bmh}^\star)$ \label{alg:BEACHES:line16}
\STATE {\bf return} $\bmh^\star$															\label{alg:BEACHES:line17}
\end{algorithmic}
\end{algorithm}

\subsection{Algorithm Simplification for Hardware Implementation} \label{sec:hw_simplification}
\revision{To enable a simpler hardware implementation of BEACHES, which is described in detail in \fref{sec:implementation}, we can approximate~$\tau_k^\star$ on  line~\ref{alg:BEACHES:line6} of \fref{alg:BEACHES} by the value~$\yshate{k}$ instead of computing the optimal value $\tau_k^\star$ exactly.} \revision{More concretely, we avoid the computations in \fref{eq:tauopt}, especially $\tau^Q_k$,  and simply use~$\yshate{k}$ in the $k$th iteration, $k = 1,2,\ldots,B$. This approximation is justified by the fact that for large values of $B$, the gap between any consecutive pair $(\yshate{k-1}, \yshate{k})$ decreases and therefore, the three values $\yshate{k-1}$, $\yshate{k}$, and $\tau_k^Q$ are typically close.}\footnote{An alternative  approach would be to replace $\tau_k ^\star$ by $\textstyle\frac{1}{2}(\yshate{k-1}+\yshate{k})$, which results in slightly higher hardware complexity but avoids evaluating SURE at the boundaries. The error-rate and MSE performance of both of these approximations is practically the same as the optimal method.}
While this approximation helps to reduce  the complexity of our hardware implementation, the simulations shown next reveal that the resulting performance is virtually indistinguishable from the original BEACHES algorithm. 
In addition, we avoid the reciprocal computations $1/B$ on line~\ref{alg:BEACHES:line7} in~\fref{alg:BEACHES} by scaling the SURE expression by $B$; we also omit the constant term $\Eo$. Both of these tricks do not affect the value of $\tau^\star$ that minimizes this expression.

%

%% file: 3d-simulation.tex

\subsection{Simulation Results}
\label{sec:SimRes}

\newcommand{\figsize}{0.43}
\begin{figure*}[tp]
\centering
\subfigure[LoS, $B=128$, $U=8$]{\includegraphics[width=\figsize\textwidth]{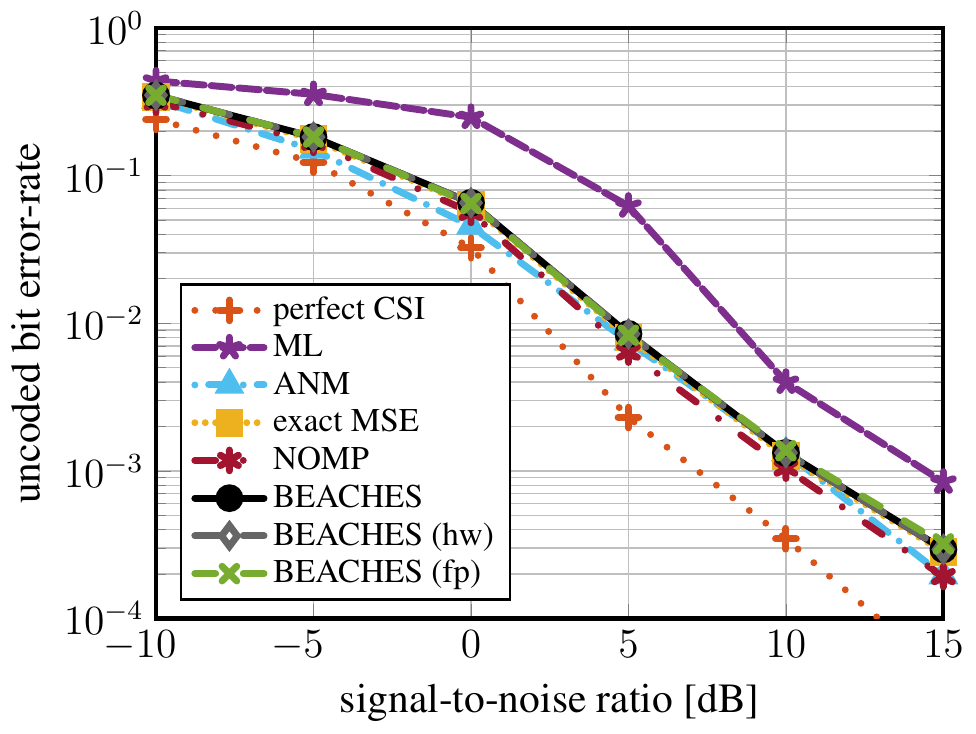}}
\hfill
\subfigure[LoS, $B=256$, $U=16$]{\includegraphics[width=\figsize\textwidth]{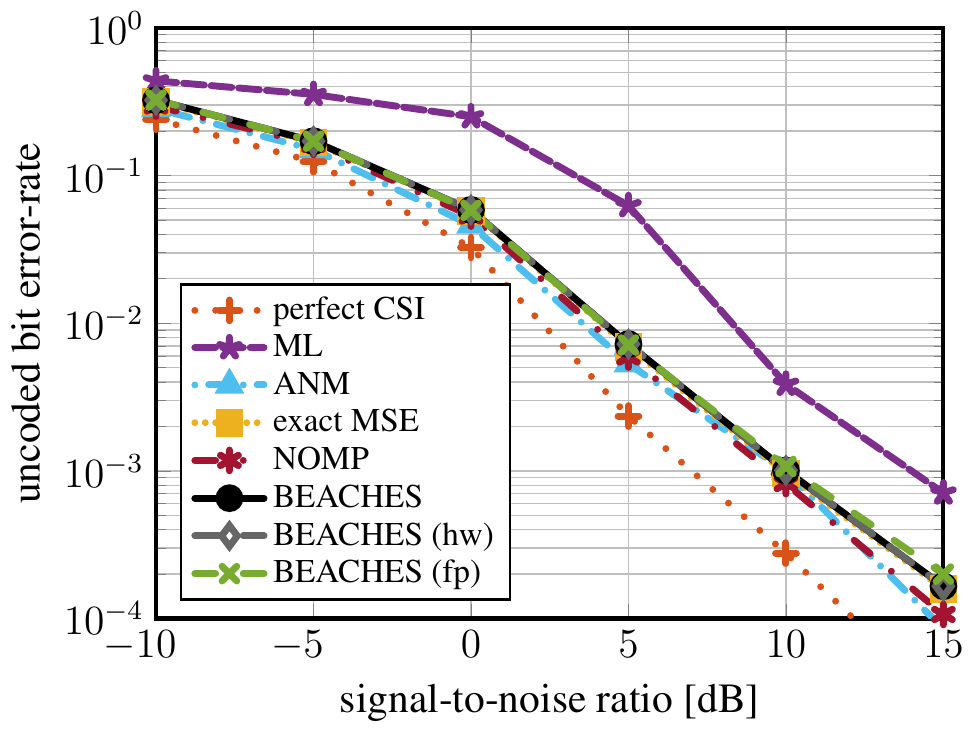}} 
\subfigure[Non-LoS, $B=128$, $U=8$]{\includegraphics[width=\figsize\textwidth]{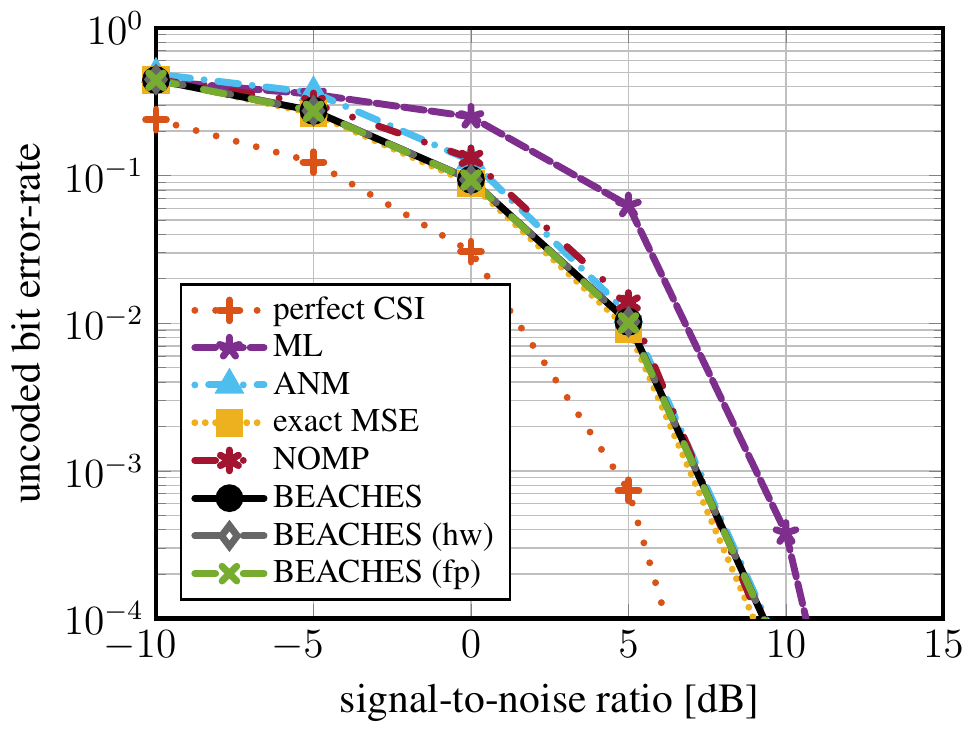}}
\hfill
\subfigure[Non-LoS, $B=256$, $U=16$]{\includegraphics[width=\figsize\textwidth]{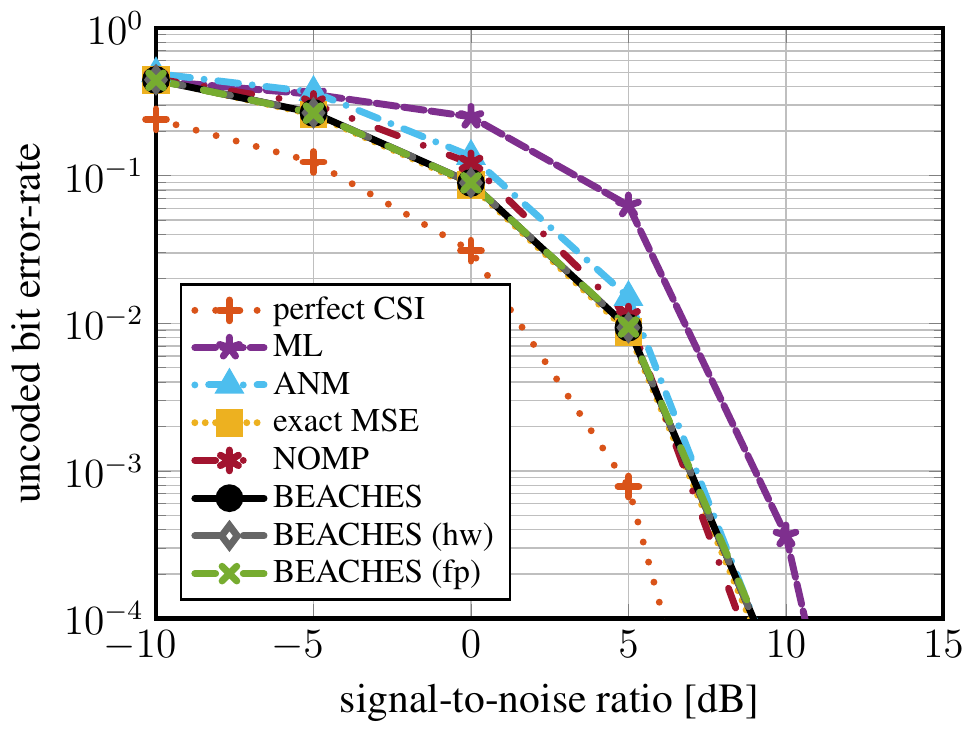}}
\caption{Uncoded bit error-rate (BER) performance of channel denoising methods for LoS and non-LoS channels. We see that BEACHES performs on par with atomic norm minimization (ANM) and Newtonized OMP (NOMP), and provides $2$\,dB to $3$\,dB SNR improvements over ML channel estimation at $\textit{BER}=10^{-3}$.}
\label{fig:BER}
\end{figure*}

\begin{figure*}[tp]
\centering
\renewcommand{\arraystretch}{1.2}	
\subfigure[LoS, $B=128$, $U=8$]{\includegraphics[width=\figsize\textwidth]{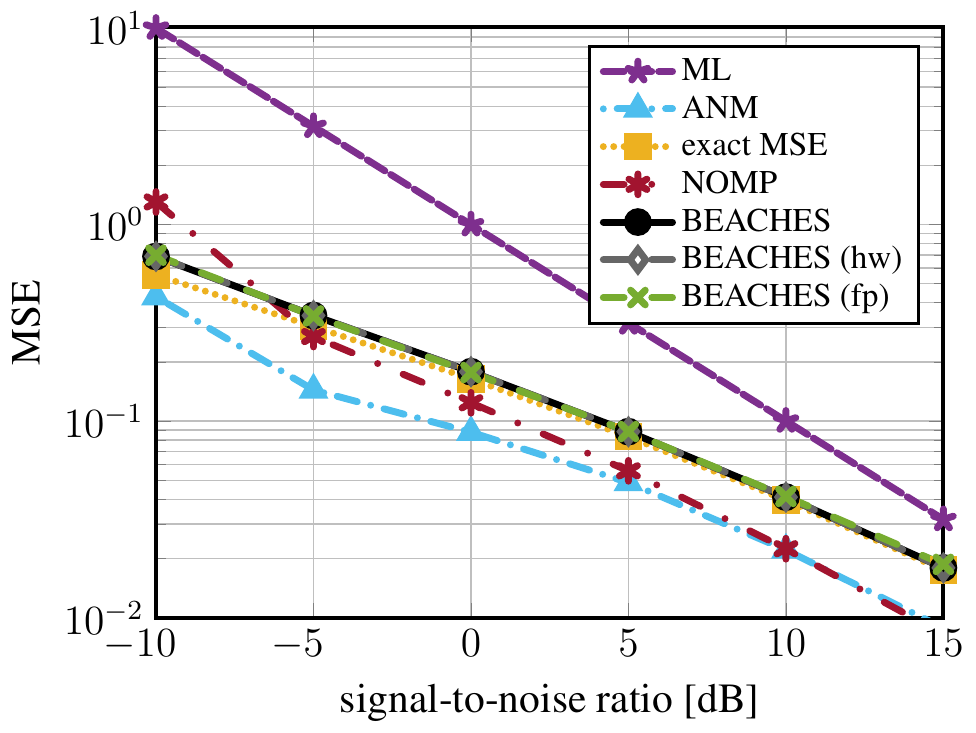}}
\hfill
\subfigure[LoS, $B=256$, $U=16$]{\includegraphics[width=\figsize\textwidth]{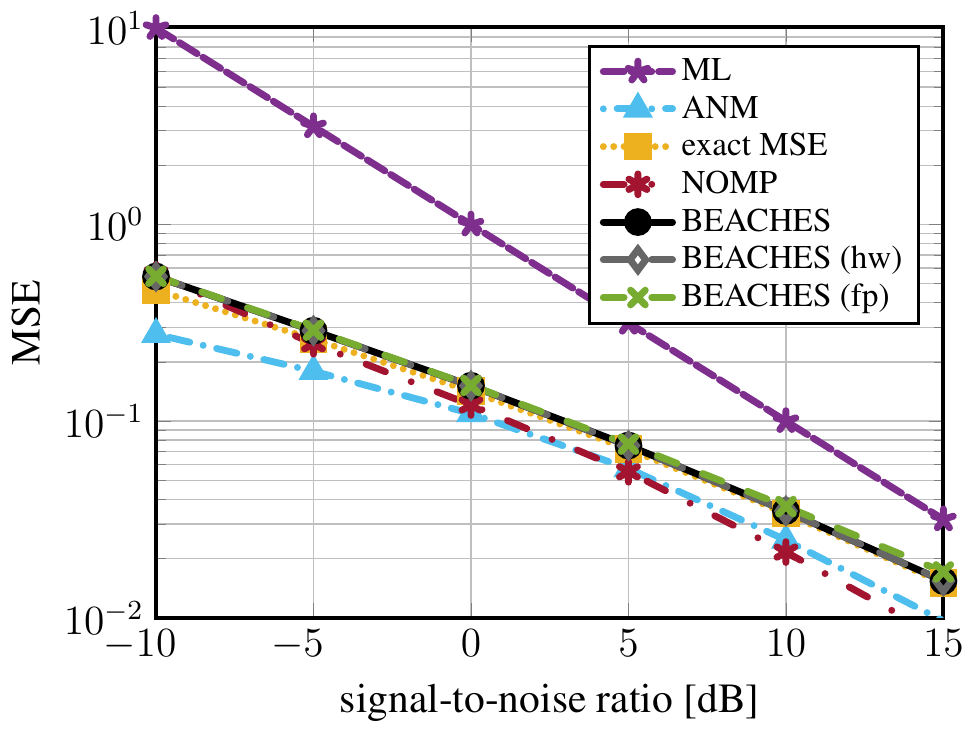}}
\subfigure[Non-LoS, $B=128$, $U=8$]{\includegraphics[width=\figsize\textwidth]{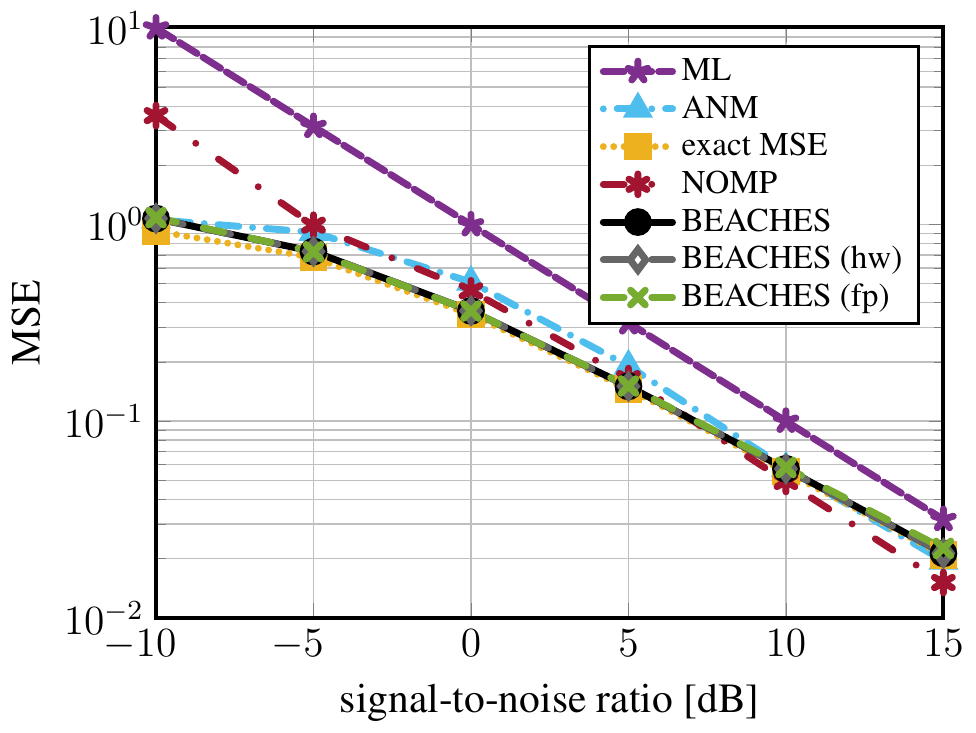}}
\hfill
\subfigure[Non-LoS, $B=256$, $U=16$]{\includegraphics[width=\figsize\textwidth]{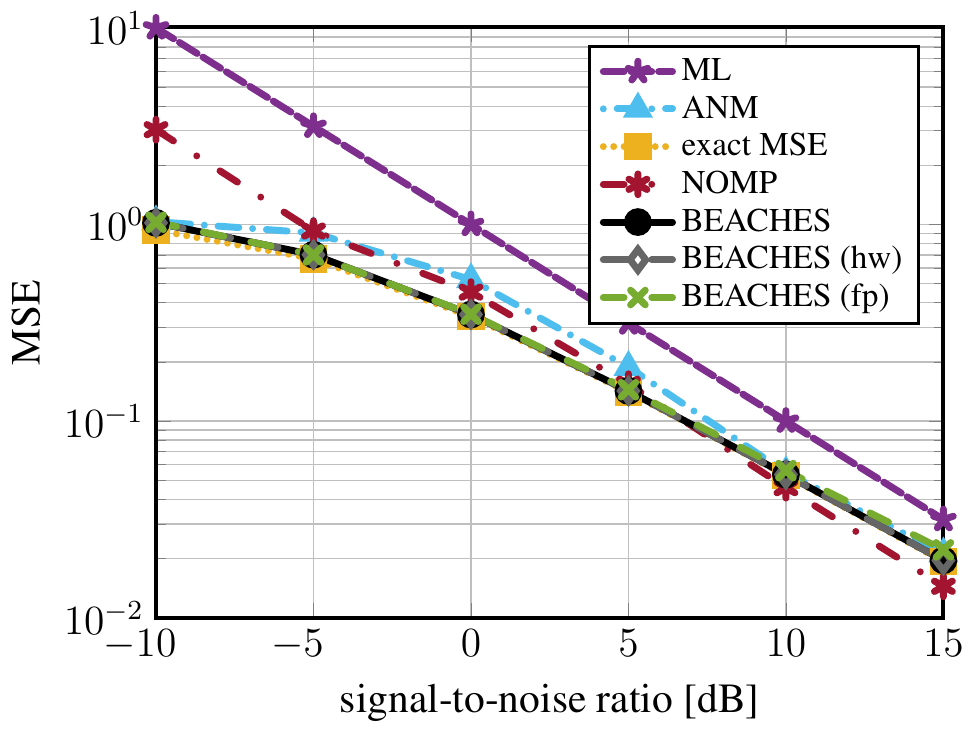}}
\caption{Mean-square error (MSE) performance of channel denoising methods for LoS and non-LoS channels. We see that BEACHES provides $2.5\times$ to $6\times$ MSE improvement over ML channel estimation at $\textit{SNR}=0$\,dB.}
\label{fig:MSE}
\end{figure*}

To demonstrate the effectiveness of BEACHES, we now present simulation results and a comparison with existing channel vector denoising methods.

\subsubsection{Simulated Scenario}
We consider a massive MU-MIMO scenario in which $U$ UEs communicate with a $B$-antenna BS over $t=1,\ldots,T$ time slots.
The input-output relation of the flat-fading system in time slot $t$ is modeled by 
\begin{align}\label{eq:BS_rec_signal}
\bmr_t=\bH \bms_t+\bmn_t.
\end{align}
Here, $\bmr_t\in\complexset^B$ is the received vector at the BS, $\bH\in\complexset^{B\times U}$ represents the (unknown) MIMO channel, 
$\bms_t = [s_{1,t}, \ldots, s_{U,t}]^\Tran$ is the transmit vector with entries chosen from a discrete constellation $\setO$ and normalized as  $\Ex{}{\|\bms_t\|^2_2} = \rho^2$, and $\bmn_t\sim\setC\setN(0,\No\bI_B)$ models thermal noise. 

During the channel estimation phase, we sequentially train each column of $\bH$ over $U$ time slots. Concretely, in each time slot $t=1,\ldots,U$, one UE is active and transmits $s_{u,t}=\rho$, whereas all others remain inactive. 
With this training scheme, the estimate of the $u$th column of the MIMO channel matrix~$\bH$ can be modeled  as $\bmy_u=\bmh_u+\bme_u$ as done in \fref{eq:antenna_model}, where the channel estimation error corresponds to $\bme\sim\setC\setN(\bZero,\Eo\bI_B)$ with variance $\Eo=\No/\rho^2$ per complex entry.
We then perform denoising independently for each column of the noisy observation of $\bH$  to obtain an improved channel matrix  $\bH^\star$.

During the data transmission phase, all UEs $u=1,\ldots,U$ transmit a constellation point from the set~$\setO$ to the BS concurrently and in the same frequency band; with the same power normalization $\Ex{}{\|\bms_t\|^2_2} = \rho^2$, as in the training phase.
\revision{Data detection uses linear minimum-mean-square-error (\mbox{L-MMSE}) equalization~\cite{studer2011asic} with the estimated matrix~$\bH^\star$.}

To characterize the performance of BEACHES and other denoising algorithms, we simulate (i) the uncoded bit error rate for 16-QAM and (ii) the channel estimation MSE as in~\fref{eq:MSE}. 
The channel matrices are generated for both a LoS and a non-LoS conditions using the QuaDRiGa mmMAGIC UMi model~\cite{QuaDRiGa_tech_rpt}, at a carrier frequency of 60\,GHz with a ULA using $\lambda/2$ antenna spacing. 
The UEs are placed randomly within a $120^\circ$ circular sector with minimum and maximum distance of $10$ and $110$ meters from the BS antenna array, respectively. In addition, we enforce a UE separation of at least $1^\circ$ (with respect to the BS antenna array) and assume optimal UE power control.

\begin{remark}
To enable the readers to perform numerical  simulations with other system parameters, channel models, or channel estimation algorithms, 
our MATLAB simulator is available at https://github.com/IIP-Group/BEACHES-simulator
\end{remark}

\subsubsection{BER Performance}
Figure~\ref{fig:BER} shows uncoded bit error rate (BER) simulation results for $B=128$ BS antennas with $U=8$ UEs, and $B=256$ BS antennas with $U=16$ UEs, for LoS and non-LoS channel conditions.\footnote{The BER at high SNR for the LoS scenario differs slightly to that of our conference paper~\cite{ghods19}, due to fewer Monte-Carlo trials in that paper.}
In addition to BEACHES as detailed in \fref{alg:BEACHES}, we show the BER of the hardware-friendly version described in \fref{sec:hw_simplification}, called ``BEACHES (hw)''  and that of our fixed-point hardware design called ``BEACHES (fp).''
We also compare our methods to the following channel estimation methods:
(i) Maximum likelihood (ML) channel estimation,
(ii) NOMP with software package provided by \cite{mamandipoor16}, where we manually tune the false alarm rate~$P_\text{fa}$ for each scenario to optimize performance,
(iii) ANM-based denoising, where we use the atomic line spectral estimation toolbox provided by~\cite{bhaskar13} (we use the exact noise variance and  the debiased output). 
As a reference, the results for ``exact MSE''  use the same soft-thresholding function as in BEACHES, but the optimal denoising parameter~$\tau^\star$ is determined by minimizing the MSE \fref{eq:MSE}, using the noiseless (ground truth) channel vector. Furthermore, ``perfect CSI'' directly uses the noiseless  channel vectors.

From \fref{fig:BER}, we see that channel vector denoising in the beamspace domain provides $2$\,dB to $3$\,dB SNR performance improvements at $\textit{BER}=10^{-3}$ compared to conventional ML channel estimation for the considered scenarios. 
\revision{The performance gains are more pronounced under LoS conditions, but significant error-rate performance improvements are also visible for non-LoS channel conditions.}
More importantly,  we observe that BEACHES performs on par with all other denoising-based channel estimation methods in terms of uncoded BER for the considered scenarios.
This observation indicates that off-the-grid denoising methods, such as NOMP and ANM, do not provide a critical performance advantage over BEACHES.
Furthermore, our hardware friendly algorithm ``BEACHES (hw)'' and the fixed-point version ``BEACHES (fp)'' deliver the same performance as BEACHES. 

\subsubsection{MSE Performance}
Figure~\ref{fig:MSE} shows the MSE of channel estimation for the same scenarios and algorithms considered in \fref{fig:BER}. 
In terms of MSE, the performance of ANM and NOMP is superior to that of BEACHES for LoS channels. We address this to the fact that the channel realizations are extremely sparse under such conditions (cf.~\fref{fig:sparse_channel}(a)). \revision{For non-LoS channels, all methods perform equally well. We address this observation to the fact that  the beamspace representation for these non-LoS channels is not sufficiently sparse (cf.~\fref{fig:sparse_channel}(b)) to leverage the off-the-grid capabilities provided by ANM and NOMP.}
These simulations also indicate that the MSE is not a particularly reliable metric to predict the BER performance of channel estimation methods in massive MU-MIMO mmWave systems.

\subsection{Complexity Scaling and Runtime Comparison}
\label{sec:complexityanalysis}

{We now compare the complexity scaling of BEACHES to that of NOMP and  ANM. We furthermore provide a MATLAB runtime comparison for LoS and non-LoS channels. 
In what follows, we assume that the complexity of a $B \times B$ matrix inversion and eigenvalue decomposition scales with~$O(B^3)$.}

\subsubsection{Complexity Scaling}
\label{sec:complexityscaling}
\revision{As mentioned in \fref{sec:BEACHES_alg}, the complexity of BEACHES scales with ${O}(B\log(B))$ and is dominated by the FFT, IFFT,  and sorting operations.}

\revision{The complexity of NOMP scales with~\cite{mamandipoor16} 
\begin{align} \label{eq:complexityNOMP}
{O}(K \gamma B\log( \gamma B) +K^2 B + BK^3 + K^4),
\end{align}
where $\gamma$ is the frequency oversampling factor (typically set to~$4$) and $K$ represents the number of NOMP iterations, which also specifies the number of detected complex sinusoids. 
The exact value of $K$ is determined internally by NOMP and depends on a number of factors, including the false alarm rate $P_\text{fa}$, the SNR, and the channel scenario, all of which affect the sparsity level of the observation vector. 
We have observed typical values for $K$ ranging from $2$ to $45$ for the simulated scenarios in \fref{sec:SimRes}.
For large $B$, the complexity of NOMP is dominated by the term $K\gamma B\log(\gamma B)$ in~\fref{eq:complexityNOMP}. Hence,  by ignoring the term $BK^3$ in~\fref{eq:complexityNOMP}, the complexity of NOMP is at least $K\gamma/3$ times higher than that of BEACHES---we confirm this observation in the runtime comparison  of \fref{sec:runtime}.}

\revision{The complexity of ANM scales with ${O}(K' B^3)$, where $K'$ is the number of iterations of the fast alternating direction method of multipliers (ADMM) implementation provided by~\cite{bhaskar13}.
Each algorithm iteration  requires a projection onto the semidefinite cone, which can be implemented via an eigenvalue decomposition whose complexity scales with ${O}(B^3)$ \cite{GV96}.
We have observed typical values of $K'$ ranging from $130$ to $360$ for  the simulated scenarios in   \fref{sec:SimRes}.
Consequently, ANM has orders-of-magnitude higher complexity than BEACHES, especially for a large number of BS antennas $B$---we confirm this observation by the runtime comparison detailed next.}

\begin{table}[tp]
	\centering
	\renewcommand{\arraystretch}{1.1}
	\begin{minipage}[c]{1\columnwidth}
		\centering
		\caption{MATLAB runtimes in milliseconds (and normalized runtimes) on an Intel core i5-7400 CPU with 16\,GB RAM.}
		\label{tbl:runtimes}
		\begin{tabular}{@{}lccc@{}}
			\toprule
			Scenario & BEACHES & NOMP & ANM \\
			\midrule
 	
			$B = 128$, LoS     & 0.57 (1$\times$)  & 28.36 (50$\times$)        & 5\,221 (9\,100$\times$)   \\
			$B = 128$, non-LoS & 0.40 (1$\times$)  & 260.4 (650$\times$)       & 7\,725 (19\,000$\times$)  \\
			\midrule
			$B = 256$, LoS     & 1.64 (1$\times$)  & 199.9 (120$\times$)       & 47\,968 (29\,000$\times$) \\ 
			$B = 256$, non-LoS & 1.45 (1$\times$)  & 2\,204 (1\,500$\times$) & 83\,750 (58\,000$\times$) \\						
			\bottomrule
		\end{tabular}
	\end{minipage}
\end{table}

\subsubsection{Runtime Comparison} \label{sec:runtime}
While the performance in terms of uncoded BER is comparable for all considered channel estimation methods, BEACHES exhibits (often significantly) lower complexity than NOMP and ANM. 
To reinforce this claim, we measured their MATLAB runtimes in milliseconds on an Intel core i5-7400 CPU with 16\,GB RAM at a signal-to-noise ratio (SNR) of $5$\,dB; at higher SNRs, the runtimes of NOMP and ANM increase by up to $2\times$ whereas the runtime of BEACHES remains unaffected. 
\fref{tbl:runtimes} demonstrates that the runtime of BEACHES is orders-of-magnitude lower than that of NOMP (up to $1\,500\times$) and ANM (up to $58\,000\times$), while the speedup is more pronounced for $B=256$ BS antennas than for $B=128$ BS antennas. 

\begin{remark}
\revision{MATLAB runtime measurements can only serve as a proxy to the true complexity as they hide the effect of coding details. Nevertheless, the extreme speedups for channel vector denoising shown in \fref{tbl:runtimes} confirm the inherent complexity advantages of BEACHES over ANM and NOMP  reflected in our analytical expressions provided in \fref{sec:complexityscaling}.}
\end{remark}

\begin{remark}
The complexity scaling analysis and runtime comparison in \fref{tbl:runtimes} hides an important aspect: NOMP and ANM can be used for compressive channel estimation whereas BEACHES can only be used for beamspace channel vector denoising. The development of efficient off-the-grid channel estimation methods specialized to beamspace channel vector denoising is an interesting open research problem.
\end{remark}

%% file: 4-implementation.tex
\section{VLSI Design and FPGA Implementation}
\label{sec:implementation}

We now describe a VLSI architecture of the simplified version of BEACHES described in \fref{sec:hw_simplification}, and present reference FPGA implementation results.

\begin{figure*}[tp]
	\centering
	\includegraphics[width=0.99\textwidth]{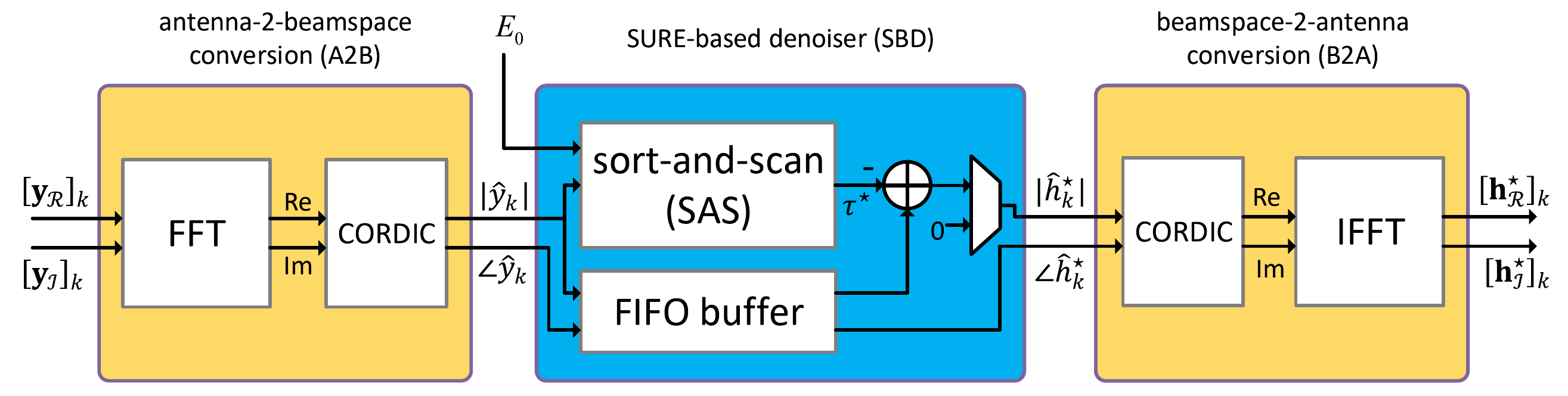}
	\vspace{-0.1cm}
	\caption{High-level VLSI architecture of BEACHES (hw). The architecture operates in input/output streaming mode and consists of three modules: an antenna-to-beamspace (A2B) conversion module, a SURE-based denoiser (SBD) module, and a beamspace-to-antenna (B2A) conversion module. The only required parameter is the variance $\Eo$ of the channel estimation noise, which is known in practical systems. }
	\label{fig:arch_overview}
\end{figure*}

\subsection{Architecture Overview} \label{sec:arch_overview}
\fref{fig:arch_overview} provides a high-level overview of the proposed VLSI architecture that implements the hardware (hw) version of BEACHES presented in \fref{sec:hw_simplification}.
The architecture consists of three main modules: (i) an antenna-to-beamspace (A2B) conversion module, (ii) a SURE-based denoiser (SBD) module, and (iii) a beamspace-to-antenna (B2A) conversion module. 
\revision{The A2B module transforms the received antenna-domain channel vector~$\bmy$ into the beamspace domain vector~$\hat{\bmy}$ as given by \fref{eq:beamspace_model}. The same module also converts the individual entries of the vector~$\hat{\bmy}$ from Cartesian coordinates to polar coordinates, which simplifies the adaptive denoising procedure.
The SBD module implements SURE-based denoising, i.e., first identifies the optimal denoising parameter~$\tau^\star$ and then applies the shrinkage to the magnitudes of the beamspace vector entries $\hat{y}_k$, $k=1,\dots, B$.
The B2A module converts the entries of the denoised beamspace vector~$\hat{h}^{\star}_k$, $k=1,\ldots,B$, from polar into Cartesian coordinates. The same module also transforms the denoised beamspace vector~$\hat{\bmh}^{\star}$ back into the antenna domain~${\bmh}^{\star}$.
To maximize throughput, the proposed architecture relies on input/output streaming. Concretely, the architecture reads a new channel vector entry and generates a new denoised entry (after a certain processing latency) in each clock cycle.}
The streaming nature of the proposed architecture also reduces control overhead and the need for additional storage of intermediate results.

\begin{figure*}[tp]
	\centering
	\includegraphics[width=0.99\textwidth]{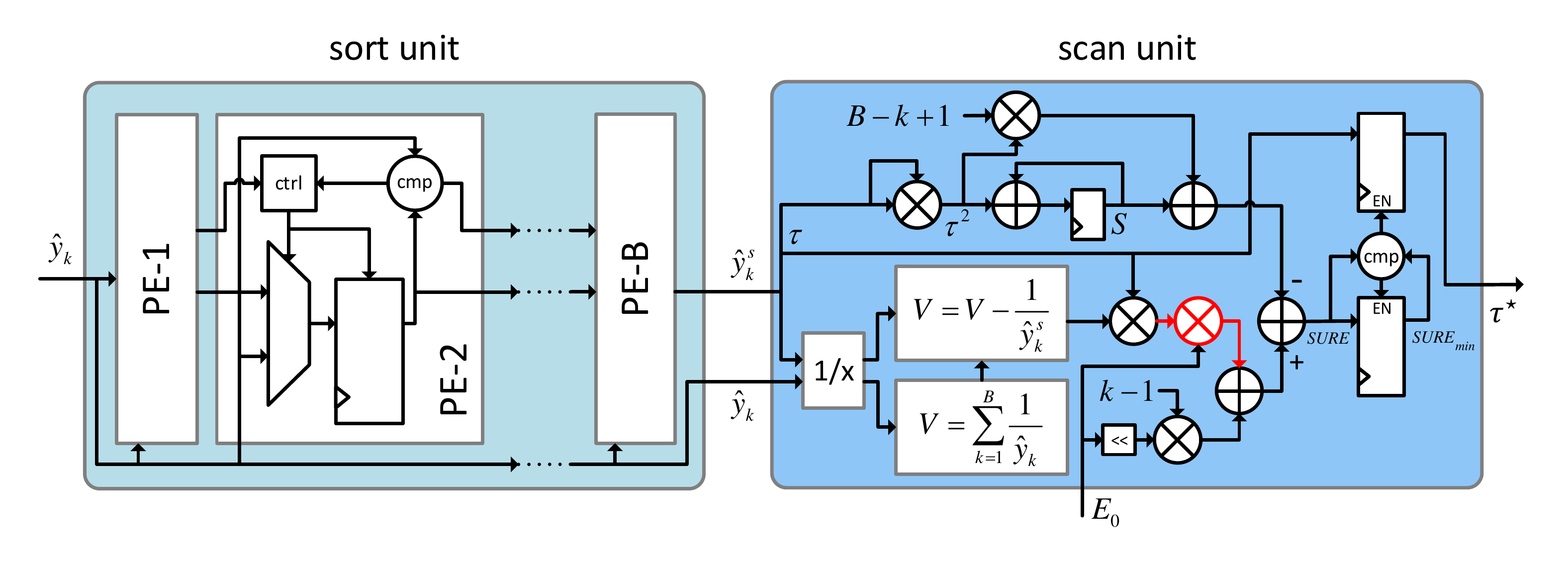}

	\vspace{-0.3cm}
	\caption{Architecture details of the sort-and-scan (SAS) module that supports I/O streaming. The sort unit (left) sorts the magnitudes of the beamspace vector in descending order (PE-$B$ contains the smallest sorted element) using a linear array of comparators; the scan unit uses the sorted outputs to determine the optimal SURE threshold $\tau^\star$ for shrinkage-based denoising. \revision{The critical path of this architecture is in the scan unit and highlighted with red color.}}
	\label{fig:arch_SAS}
\end{figure*}

\subsection{Architecture Details} \label{sec:arch_details}

The architecture details of the three modules shown in \fref{fig:arch_overview} are as follows.

\subsubsection{Antenna-to-Beamspace (A2B) Conversion Module} \label{sec:A2B}
\revision{As shown in \fref{fig:arch_overview}, the A2B conversion module contains a streaming FFT that transforms the noisy antenna-domain channel vector into the beamspace domain.} In our implementation, we use a Xilinx LogiCORE FFT IP with radix-2 pipelined I/O streaming, which reads and generates one vector entry per clock cycle.
 As a consequence, the FFT core completes computation of the $B\times 1$ beamspace vector $\hat{\bmy}$ every $B$ clock cycles. To reduce area, we configured the FFT core to scale down the intermediate values by a factor of two in each of the $\log_2(B)$ FFT stages. 
 This configuration reduces the dynamic range in each FFT stage and also reduces resource utilization. 
Additionally, this scaling approach yields FFT outputs that have smaller dynamic range compared to the unscaled case, which allows for more compact fixed-point data representation and reduces storage requirements in the subsequent modules.

\revision{After FFT processing, each complex-valued beamspace domain sample $\hat{y}_k$  is passed through a vectoring CORDIC, which converts the Cartesian number representation into polar coordinates. This transform simplifies the soft-thresholding operation, as it only needs to be applied to the magnitude of each entry in the SBD module---the phase remains untouched.} The CORDIC is  implemented using a Xilinx LogiCORE IP. The number of microrotations in the CORDIC core is determined by the IP so that the achieved accuracy is 10\,bit; see \fref{sec:fp_details} for more details on the  fixed-point parameters of our design. 	

\subsubsection{SURE-based Denoiser (SBD) Module}
As shown in \fref{fig:arch_overview}, this module consists of a first-in first-out (FIFO) buffer, a module to perform sort-and-scan (SAS) in order to determine the optimal denoising threshold, and logic (a subtractor and a multiplexer) to apply soft-thresholding to the values in the FIFO buffer. 
The role of the FIFO buffer is to delay the inputs of the SBD module so that they are ready as soon as the optimal threshold $\tau^\star$ has been computed. 
{The FIFO buffer has a depth of $2B + 5$ entries, corresponding to the latency of the SAS submodule as detailed in the next paragraphs.} 
The details of the SAS architecture are shown in \fref{fig:arch_SAS}.
The architecture  consists of a sort unit and a subsequent scan unit, corresponding to lines~\ref{alg:BEACHES:line4} to~\ref{alg:BEACHES:line14} of \fref{alg:BEACHES}. The following paragraphs summarize the most important architecture details.

As depicted in \fref{fig:arch_SAS}, the sort unit consists of an array of~$B$ identical processing elements (PEs). 
The details of the PEs are shown for the second PE (PE-$2$), which consists of (i) a register to keep one of the sorted elements, (ii) a multiplexer that selects whether the new input data or the value stored in the previous PE should enter the register,  (iii) a comparator (denoted by ``cmp'') that compares the new input value with the value stored in the PE's register, and (iv) a control unit (denoted by ``ctrl'') that determines the multiplexer output and whether the register must be updated.
As for the FFT core, the sort unit is using I/O streaming, i.e., the architecture continuously reads and generates data. 
\revision{This architecture also allows for a seamless integration with the scan unit (discussed below), and eliminates the need to buffer the sorted data separately in a memory for the scan unit to work on.}
\revision{The sort unit sorts the data as they enter, by finding the appropriate position within the array for each new input data, similar to an insertion sort algorithm.} Assume that~$k$ entries of a $B\times 1$ vector have already been sorted and reside in the PEs $1$ to $k$. In the next clock cycle, the $(k+1)$th (unsorted) element enters the sort unit and is broadcast to all PEs. Each PE compares the new element with the value stored in its own register, and additionally, receives the result of the same comparison from its preceding PE. 
For the case of sorting in descending order (i.e., PE-$B$ stores the smallest element), the new input data will be placed in PE-$m$, if the new element is larger than the data stored in PEs $1, \ldots, m-1$ and smaller than or equal to the value stored in PE-$m$. At the same time, the PEs $m, \ldots, k$ will pass their previously stored values to their adjacent PE (e.g., PE-$m$ to PE-$(m+1)$), so that no data is lost. \revision{This approach is repeated until all $B$ elements are sorted in all of the PEs at the clock cycle after receiving the last element.}
When loading the first element of the next denoising problem, PE-$B$ will pass its value (which is the smallest element of the last channel vector) to the scan unit and will receive the data of PE-$(B-1)$, and therefore the sorted data will be flushed at the same time the next problem is being loaded and sorted.

\begin{remark}
Although the algorithm complexity of BEACHES is ${O}(B\log(B))$, the implemented sorting architecture has a hardware complexity of ${O}(B^2)$ in terms of the area-delay product. The reason for this architecture choice is the fact that this sorting method supports I/O streaming without a significant overhead in terms of latency and buffering. Furthermore, our implementation results in \fref{sec:fpgaresults} demonstrate that this architecture is efficient for the targeted BS antenna numbers. 
\end{remark}

\revision{The scan unit is depicted in \fref{fig:arch_SAS}. 
In order to initialize the cumulative sum of reciprocals denoted by $V$ on line \ref{alg:BEACHES:line2} of \fref{alg:BEACHES}, the scan unit receives the entries of $\yhat$  at the same time they enter the sort unit. The reciprocal values of the entries of $\yhat$ are computed sequentially using a look-up-table (LUT) with $512$ entries and are accumulated in a register. Therefore, the cumulative sum of reciprocals is ready once the scan unit receives the the last element of $\yhat$. At the same time, the first sorted entry of $\yhat$ comes out of the sort unit. As the scan unit receives the sorted elements $\hat{y}^s_k$, it updates the value of the quantity $V$ according to the line \ref{alg:BEACHES:line13} of \fref{alg:BEACHES}. The rest of the scan unit contains adders/subtractors and multipliers to compute SURE corresponding to line \ref{alg:BEACHES:line7} of the \fref{alg:BEACHES} (with modifications detailed in \fref{sec:hw_simplification}). Finally, the registers and the comparator at the right end of the scan unit in \fref{fig:arch_SAS}, implement the conditional assignments corresponding to the lines \ref{alg:BEACHES:line8} to \ref{alg:BEACHES:line11} of the algorithm.}

\revision{The critical path of the proposed BEACHES architecture is in the scan unit as indicated with red color in \fref{fig:arch_SAS}. The critical path originates in a pipeline flip-flop, goes through a real-valued multiplier, and ends in another pipeline flip-flop. For the sake of simplicity, the pipeline registers are not shown.}

\subsubsection{Beamspace-to-Antenna (B2A) Conversion Module}
As shown in \fref{fig:arch_overview}, the B2A conversion module resembles that of the A2B module.
This module contains a rotation CORDIC, implemented by Xilinx LogiCORE IP, to transform the denoised entries from polar into Cartesian coordinates, and a Xilinx LogiCORE FFT IP to convert the denoised beamspace entries into the antenna domain.
The FFT core is configured to perform an IFFT without scaling in any of its stages. The unscaled configuration results in a word-length growth in every stage. However, since the beamspace domain signals are already scaled by the FFT core in the A2B module, the same word-length as the input channel entries is sufficient to accommodate the dynamic range of the outputs from the unscaled IFFT.

\begin{table*}[tp]
\centering
\renewcommand{\arraystretch}{1.1}
\begin{minipage}[c]{1\textwidth}
\vspace{-0.1cm}
    \centering
    \caption{Implementation results for different numbers of BS antennas $B$ on a Xilinx Virtex-7 XC7VX690T FPGA.}
       \label{tbl:implresults}
  \begin{tabular}{@{}lcccc@{}}
  \toprule
  BS antennas $B$ & $64$ & $128$ & $256$ & $512$\\
  \midrule
  {Slices} & 1\,532  (1.41\%) & 2099  (1.94\%) & 3\,089  (2.85\%) & 4\,886  (4.51\%) \\
  {LUTs} & 4\,564 (1.05\%) & 6\,391 (1.48\%) & 9\,394 (2.17\%) & 14\,449 (3.34\%)\\
  -- {logic LUTs} & 3\,970 (0.92\%) & 5\,566 (1.28\%)& 8\,336 (1.92\%) & 13\,523 (3.12\%) \\
  -- {memory LUTs} & 594 (0.34\%)& 825 (0.47\%) & 1\,058 (0.61\%) & 926 (0.53\%)\\
  {Flipflops} & 5\,561 (0.64\%) & 7\,015 (0.81\%)& 9\,282 (1.07\%) & 13\,133 (1.52\%) \\
  {DSP48 units} & 24 (0.67\%) & 32 (0.89\%) & 32 (0.89\%) & 40 (1.11\%)\\
  {Block RAMs} & 1 (0.07\%) & 1 (0.07\%) & 2 (0.14\%) & 5.5 (0.37\%) \\
  \midrule
 Max.\ clock frequency [MHz] & 303 & 303 & 303 & 294 \\
  {Latency [clock cycles]} & 575 & 972 & 1752 & 3301 \\
  \revision{Latency [$\mu$s]} & 1.8 & 3.2 & 5.8 & 10.9 \\
  {Throughput\footnote{The throughput is given in million vectors denoised per second and calculated as $f/B$, where $f$ is the maximum clock frequency.} [Mvectors/s]} & 4.73 & 2.36 & 1.18 & 0.57 \\
  {Power consumption\footnote{Statistical power estimation at maximum clock frequency and for 1.0\,V supply voltage.} [W]} & 0.76 & 0.87 & 0.98 & 1.29 \\
  \midrule
  {Efficiency [Mentries/s/LUT]}  & {66\,389} & {47\,410} & {32\,255} & {20\,970} 
  \tabularnewline  
  \bottomrule
  \end{tabular}
  \end{minipage}
  \end{table*}

\begin{table*}[tp]
	\centering
	\renewcommand{\arraystretch}{1.1}
	\begin{minipage}[c]{1\textwidth}
		\centering
		\caption{FPGA resource and latency breakdown for different numbers of BS antennas $B$  on a Xilinx Virtex-7 XC7VX690T FPGA.}
		\label{tbl:Breakdown}
		\begin{tabular}{@{}lccccccccccccccc@{}}
			\toprule
			BS antennas $B$ & \multicolumn{3}{c}{64} && \multicolumn{3}{c}{128} && \multicolumn{3}{c}{256}  && \multicolumn{3}{c}{512}\\
			\cmidrule{2-4}  \cmidrule{6-8} \cmidrule{10-12} \cmidrule{14-16}
			{Module} & A2B & SBD & B2A  &&  A2B & SBD & B2A &&  A2B & SBD & B2A &&  A2B & SBD & B2A\\
			\midrule
			{LUTs} & 1\,650 & 1\,517 & 1\,408             &&   1\,947 & 2\,798 & 1\,658   && 2\,176 & 5\,331 & 1\,899    &&   2\,381 & 9\,985 & 2\,092 \\
			-- {logic LUTs } & 1\,354 & 1\,450 & 1\,177   &&   1\,542 & 2\,691 & 1\,345   && 1\,701 & 5\,144 & 1\,503    &&   1\,885 & 9\,978 & 1\,669 \\
			-- {memory LUTs} & 296 & 67 & 231        &&   405 & 107 & 313      && 475 & 187 & 396       &&   496 & 7 & 423 \\
			{Flipflops} & 2\,470 & 994 & 2\,097           &&   2\,834 & 1\,769 & 2\,412   && 3190 & 3312 & 2780    &&   3\,646 & 6\,374 & 3\,113 \\
			{DSP48 units} & 9 & 5 & 10                  &&    13 & 5 & 14         && 13 & 5 & 14           &&   17 & 5 & 18 \\  
			\midrule
			\revision{Latency [clock cycles]} & 218 & 136 & 221                  &&    353 & 264 & 355         && 615 & 520 & 617           &&   1134 & 1032 & 1135 \\  
			\bottomrule
		\end{tabular}
	\end{minipage}
\end{table*}

\subsection{Fixed-Point Parameters}  \label{sec:fp_details}
To maximize hardware-efficiency, we use two's complement fixed-point arithmetic. 
The number of bits used for signals in our implementation has been determined based on extensive bit error-rate (BER) simulations, with the goal of achieving near-floating-point performance while minimizing area.
\revision{For the antenna domain channel entries, we use $16$ bits of which are $8$ fractional bits. Due to the FFT scaling described in \fref{sec:A2B}, $10$ bits are sufficient for the beamspace vector entries. Therefore, the entries of the vectors $\hat{\bmy}$ and $\hat{\bmy}^{\star}$ are represented with $10$ bits of which are $8$ fractional bits, in both the Cartesian and polar coordinates.}
\revision{Since the SBD module (shown in~\fref{fig:arch_overview}) operates in the beamspace domain, most of its signals are represented with $10$-bit numbers. For the quantity $E_0/B$, we use $16$-bit numbers with $15$ fraction bits. We have eliminated the term $E_0$ in line \ref{alg:BEACHES:line8} of \fref{alg:BEACHES}, since it is a constant and does not affect the value of the optimal threshold. For the entries of the LUT which is used to compute reciprocals in the scan unit, we use $12$-bit numbers with $2$ fraction bits.  Other intermediate signals in the scan unit have customized word-lengths to accommodate temporary dynamic range growth caused by multiplications and additions.}

\revision{The BER and MSE performance of our fixed-point BEACHES architecture are shown in \fref{fig:BER} and \fref{fig:MSE}, respectively, where ``fp'' stands for fixed-point performance. Clearly, the loss due to finite-precision arithmetic is negligible compared to the reference floating-point MATLAB model.}

\subsection{FPGA Implementation Results} \label{sec:fpgaresults}

To demonstrate the efficacy of BEACHES in practice, we have implemented our architecture on a Xilinx \mbox{Virtex-7} XC7VX690T FPGA (speed grade $-3$) for various BS antenna configurations  ($B=64,128,256,512$). The implementation results are summarized in \fref{tbl:implresults}, and confirm the low complexity of BEACHES when implemented in hardware. 
In fact, the resource utilization (in terms of slices, LUTs, flip-flops, DSP48 units, and block RAMs) is within a few percent of the total FPGA resources.
Furthermore, we observe that the resource utilization (measured in terms of LUTs and flip-flops) increase roughly linearly with the number of BS antennas, which is mainly due to the fact that the number of comparison PEs in the SAS module grows linearly in $B$. 
Similarly, we see that the throughput  (measured in million vectors denoised per second) decreases roughly linearly in~$B$. 
 The hardware efficiency (measured in million entries per LUT) also reduces approximately linearly in $B$, which is intuitive as more work must be carried out by BEACHES for systems with more BS antennas. 
\fref{tbl:Breakdown} shows a detailed area breakdown of our FPGA designs. We can see that for $B=64$, the three modules (A2B, SBD, and B2A) occupy about the same amount of resources. However, when increasing $B$, we see that the complexity of the SBD unit dominates. This is due to the fact that the complexity of the sorting module is the only one whose resources grow linearly in $B$. Evidently, if one is interested in further increasing the number of BS antennas~$B$, alternative sorting engines should be used. 

\begin{remark}
\revision{The BEACHES design supporting $B = 512$ BS antennas, which achieves the lowest throughput in \fref{tbl:implresults}, denoises up to $570\,000$ channel vectors per second. By assuming a system with $U = 16$ UEs, this architecture can denoise up to $35\,625$ channel matrices per second, i.e., one channel matrix every $28$\,\text{$\mu$s}. Since typical coherence times of mmWave channels are in the order of several milliseconds \cite{he18}, channels need to be estimated roughly once every $1000$\,\text{$\mu$s}. Therefore, the throughput of our FPGA designs are well above what is required for mmWave channel estimation.}
\end{remark}

\revision{We conclude by noting that there exists, to the best of our knowledge, no channel vector denoising implementation in the open literature that would enable a fair comparison.
Nevertheless, a handful of results in the literature are concerned with hardware designs for sparsity-based channel estimation algorithms, such as \cite{MaechlerThesis, korrai18, birke18}. The hardware designs reported in \cite{MaechlerThesis} are for wideband single-input single-output (SISO) channels in 3GPP-LTE systems. These implementations exploit sparsity in the delay domain and are based on three serial greedy pursuit algorithms, namely matching pursuit, gradient pursuit, and OMP. 
The FPGA design reported in~\cite{korrai18} focuses on channel estimation of indoor SISO systems---again, this result exploits sparsity in the delay domain.  The results in~\cite{birke18} focus on short-range, point-to-point, indoor communication with hybrid precoding---a direct comparison of these methods to our work is difficult.
We reiterate that all these results do not focus on massive MU-MIMO mmWave denoising in the beamspace domain and require the user to set certain algorithm parameters. In contrast, BEACHES is specialized to perform adaptive denoising in the beamspace domain while only requiring knowledge of the noise variance.}

%% file: 5-conclusions.tex

\section{Conclusions}
\label{sec:conclusions}
We have proposed a nonparametric channel estimation  algorithm for massive MU-MIMO mmWave systems, which we call BEAmspace CHannel EStimation (BEACHES).
BEACHES exploits channel sparsity of mmWave channels in the beamspace domain in order to perform adaptive denoising via Stein's unbiased risk estimate (SURE).
We have established that BEACHES achieves MSE-optimal performance in the large-antenna limit.
For realistic LoS and non-LoS mmWave channel models, we have shown that BEACHES performs on par with sophisticated channel estimation algorithms in terms of uncoded bit-rate performance but at orders-of-magnitude lower complexity. 
As a direct consequence of the nonparametric nature of our algorithm, BEACHES continues to minimize the channel estimation MSE even in scenarios where no sparsity can be exploited (e.g., for Rayleigh fading channels). 

In order to demonstrate the practicality of BEACHES, we have developed reference FPGA implementations for massive MU-MIMO mmWave systems with hundreds of BS antennas. 
Our results are a proof-of-concept that high-quality mmWave channel estimation can be performed at high throughput and in a hardware-efficient manner. 

There are many avenues for future work. \revision{An adaptation of BEACHES to single-carrier (SC) transmission in mmWave channels is a challenging open research problem.}
\revision{The development of nonparametric channel estimation methods that do not need knowledge of the noise variance is part of ongoing work.}
\revision{An extension of BEACHES to basestation architectures that use decentralized baseband processing \cite{li17d} to reduce interconnect bottlenecks is an interesting open research problem.}
Finally, alternative sorting architectures might be necessary when targeting systems with thousands of antenna elements.

%% file: 6-MSEtoSURE.tex
\section{Proof of \fref{thm:MSEappox}}
\label{app:MSEtoSURE}
We first derive the general form for SURE with complex-valued signals. The MSE for a weakly-differentiable estimator function {$\mu(\yhat)$} is defined as
\begin{align}
\textit{MSE} = \textstyle\frac{1}{B}\Ex{}{\|\mu(\yhat)-\hhat\|_2^2}\!.
\end{align}
Note that expectation is with respect to the noisy observation~$\yhat$. 
We decompose the complex-valued vector $\yhat$ into the real part $\yhat_{\mathcal{R}} \sim \setN(\bmy_{\mathcal{R}};\hat{\bmh}_{\mathcal{R}},\textstyle\frac{\Eo}{2} \bI_B)$ and imaginary part $\yhat_{\mathcal{I}} \sim \setN(\bmy_{\mathcal{I}};\hat{\bmh}_{\mathcal{I}},\textstyle\frac{\Eo}{2} \bI_B)$ and define $g(\yhat)=\mu(\yhat)-\yhat$.
Hence,
\begin{align}\label{eq:MSE_complex}
\textit{MSE}=&\, \textstyle \frac{1}{B}\Ex{}{\|g(\yhat)+\yhat-\hhat\|_2^2}  \\  
= &\, \textstyle\frac{1}{B}\Ex{}{\|g(\yhat)\|_2^2} + \frac{1}{B}\Ex{}{\|\yhat-\hhat\|_2^2} \nonumber \\ 
&\, + \textstyle \frac{1}{B}\Ex{}{2\left[g(\yhat)^\Herm(\yhat-\hhat)\right]_{\mathcal{R}}} \!. \label{eq:longMSEexpression}
\end{align}
The last term can be 
{expanded} as follows:
\begin{align}
& \textstyle\frac{2}{B}\Ex{}{\left[g(\yhat)^\Herm(\yhat-\hhat)\right]_{\mathcal{R}}}   \\
&   = \textstyle\frac{2}{B}\Ex{}{{g_{\mathcal{R}}(\yhat)^\Tran(\yhat_{\mathcal{R}}-\hhat_{\mathcal{R}})}} 
+\textstyle\frac{2}{B}\Ex{}{{g_{\mathcal{I}}(\yhat)^\Tran(\yhat_{\mathcal{I}}-\hhat_{\mathcal{I}})}}\!.\notag
\end{align}
We can now expand $\textstyle\frac{2}{B}\Ex{}{{g_{\mathcal{R}}(\yhat)^\Tran(\yhat_{\mathcal{R}}-\hhat_{\mathcal{R}})}}$, which yields
\begin{align}
\textstyle\frac{2}{B}&\Ex{}{{g_{\mathcal{R}}(\yhat)^\Tran(\yhat_{\mathcal{R}}-\hhat_{\mathcal{R}})}} \nonumber \\ 
=&\, \textstyle\frac{2}{B}\int_{\yhat} f^{\setC\setN}\left(\yhat;\hhat,{\Eo} \bI_B\right) \sum_{b=1}^{B} [g_{\mathcal{R}} (\yhat)]_b \times \nonumber \\
&\, ([\yhat_{\mathcal{R}}]_b-[\hhat_{\mathcal{R}}]_b) d \yhat  \\ 
= &\, \textstyle\frac{2}{B} \int_{\yhat_{\mathcal{I}}} f^{\setN}\left(\yhat_{\mathcal{I}};\hhat_\mathcal{I},\frac{\Eo}{2} \bI_B\right) \sum_{b=1}^{B} \int_{\revision{[\yhat_{\mathcal{R}}]_b}} \frac{1}{\left(2 \pi \frac{\Eo}{2}\right)^{B/2}} \times \nonumber \\
&\, \textstyle \exp{\left(-\frac{([\yhat_{\mathcal{R}}]_b-[\hhat_{\mathcal{R}}]_b)^2}{2\frac{\Eo}{2}}\right)} [g_{\mathcal{R}}(\yhat)]_b \times \nonumber\\ 
&\, ([\yhat_{\mathcal{R}}]_b-[\hhat_{\mathcal{R}}]_b) d [\yhat_{\mathcal{R}}]_b d \yhat_{\mathcal{I}} \\
\stackrel{\text{(a)}}{=} &\, \textstyle{\frac{2}{B}} \int_{\yhat_{\mathcal{I}}} f^{\setN}\left(\yhat_{\mathcal{I}};\hhat_\mathcal{I},\frac{\Eo}{2} \bI_B\right) \sum_{b=1}^{B} \int_{\revision{[\yhat_{\mathcal{R}}]_b}} \frac{1}{\left(2 \pi \frac{\Eo}{2}\right)^{B/2}} \times\nonumber  \\
&\, \textstyle \exp{\left(-\frac{(\revision{[\yhat_{\mathcal{R}}]_b}-\revision{[\hhat_{\mathcal{R}}]_b})^2}{2\frac{\Eo}{2}}\right)}  \frac{\Eo}{2} \frac{\partial {[g_{\mathcal{R}}(\yhat)]_b}}{\partial [\yhat_{\mathcal{R}}]_b} d \revision{[\yhat_{\mathcal{R}}]_b} d \yhat_{\mathcal{I}}  \\
= &\, \textstyle\frac{2}{B} \int_{\yhat_{\mathcal{I}}} f^{\setN}\left(\yhat_{\mathcal{I}};\hhat_\mathcal{I},\frac{\Eo}{2} \bI_B\right) \sum_{b=1}^{B} \int_{\yhat_{\mathcal{R}}} \frac{1}{\left(2 \pi \frac{\Eo}{2}\right)^{B/2}} \times\nonumber  \\
&\, \textstyle \exp{\left(-\frac{\|\yhat_{\mathcal{R}}-\hhat_{\mathcal{R}}\|^2}{2\frac{\Eo}{2}}\right)}  \frac{\Eo}{2} \frac{\partial {[g_{\mathcal{R}}(\yhat)]_b}}{\partial [\yhat_{\mathcal{R}}]_b} d \yhat_{\mathcal{R}} d \yhat_{\mathcal{I}}  \\
=&\, \textstyle\frac{\Eo}{B} \Ex{}{ \sum_{b=1}^{B} \frac{\partial {[g_{\mathcal{R}}(\yhat)]_b}}{\partial [\yhat_{\mathcal{R}}]_b}},  \label{eq:ReRe}
\end{align}
where step $\text{(a)}$ follows from integration by parts. Analogously, we have
\begin{align}\label{eq:ImIm}
\textstyle\frac{2}{B}\Ex{}{{g_{\mathcal{I}}(\yhat)^\Tran(\yhat_{\mathcal{I}}-\hhat_{\mathcal{I}})}} 
= \textstyle\frac{\Eo}{B} \Ex{}{ \sum_{b=1}^{B} \left(\frac{\partial {[g_{\mathcal{I}}(\yhat)]_b}}{\partial [\yhat_{\mathcal{I}}]_b}\right)}.
\end{align}
By remembering that $g(\yhat)=\mu(\yhat)-\yhat$, and replacing \fref{eq:ReRe} and~\fref{eq:ImIm} in the original MSE expression in~\fref{eq:longMSEexpression}, we obtain
\begin{align}
\textit{MSE}=&\, \textstyle\frac{1}{B}\Ex{}{\|\mu(\yhat)-\yhat\|_2^2}+\textstyle\frac{1}{B}\Ex{}{\|\yhat-\hhat\|_2^2} \nonumber \\ 
&\, +\textstyle\frac{\Eo}{B} \Ex{}{ \sum_{b=1}^{B} \left( \frac{\partial {[\mu_{\mathcal{R}}(\yhat)]_b}}{\partial [\yhat_{\mathcal{R}}]_b}+\frac{\partial {[\mu_{\mathcal{I}}(\yhat)]_b}}{\partial [\yhat_{\mathcal{I}}]_b}-2\right)}\!\!.
\end{align}
{The second term in the MSE expression above equals $\Eo$. For the first and the third terms, we omit the expectation operators to arrive at the following SURE expression:} 
\begin{align} 
\textit{SURE}=&\, \textstyle\frac{1}{B}\|\mu(\yhat)-\yhat\|_2^2+\Eo \nonumber \\ 
&\, +\textstyle\frac{\Eo}{B} \sum_{b=1}^{B} \left( \frac{\partial {[\mu_{\mathcal{R}}(\yhat)]_b}}{\partial [\yhat_{\mathcal{R}}]_b}+\frac{\partial {[\mu_{\mathcal{I}}(\yhat)]_b}}{\partial [\yhat_{\mathcal{I}}]_b}-2\right)\!,
\end{align}
for which the relationship $\Ex{}{\textit{SURE}} = \textit{MSE}$ holds.
%

%% file: 7-shrinkageSURE.tex
\section{Proof of \fref{cor:surecoolandcrazy}}
\label{app:shrinkageSURE}
In the complex domain, the soft-thresholding function has the following form~\cite[App.~A]{maleki13}:
\begin{align}
\textstyle [\eta(\hat{\bmy},\tau)]_b=\frac{\hat{y}_b}{|\hat{y}_b|}\max{\{|\hat{y}_b|-\tau,0\}},
\end{align}
where we define ${\hat{y}_b}/{|\hat{y}_b|}=0$ for $\hat{y}_b=0$.
%
%
In order to compute SURE for this shrinkage function, we will first compute its derivative of real and imaginary parts.
For $|\hat{y}_b|<\tau$, we have
\begin{align} \label{eq:mu_0}
\textstyle \frac{\partial {[\eta_{\mathcal{R}}(\hat{\bmy},\tau)]_b}}{\partial [\hat{\bmy}_{\mathcal{R}}]_b}=\frac{\partial {[\eta_{\mathcal{I}}(\hat{\bmy},\tau)]_b}}{\partial [\hat{\bmy}_{\mathcal{I}}]_b}=0.
\end{align}
For $|\hat{y}_b|>\tau$, we have
\begin{align} \label{eq:mu_r}
\textstyle \frac{\partial {[\eta_{\mathcal{R}}(\hat{\bmy},\tau)]_b}}{\partial [\hat{\bmy}_{\mathcal{R}}]_b}  =  
&\, \textstyle \frac{\partial }{\partial [\hat{\bmy}_{\mathcal{R}}]_b} \left[[\hat{\bmy}_{\mathcal{R}}]_b-\frac{\tau[\hat{\bmy}_{\mathcal{R}}]_b }{\sqrt{[\hat{\bmy}_{\mathcal{R}}]_b^2+[\hat{\bmy}_{\mathcal{I}}]_b^2}}\right] \nonumber \\ 
= &\, \textstyle 1-\tau \frac{[\hat{\bmy}_{\mathcal{I}}]_b^2}{{([\hat{\bmy}_{\mathcal{R}}]_b^2+[\hat{\bmy}_{\mathcal{I}}]_b^2)}^{3/2}}
\end{align}
and
\begin{align} \label{eq:mu_i}
\textstyle \frac{\partial {[\eta_{\mathcal{I}}(\hat{\bmy},\tau)]_b}}{\partial [\hat{\bmy}_{\mathcal{I}}]_b} = 
&\, \textstyle \frac{\partial }{\partial [\hat{\bmy}_{\mathcal{I}}]_b} \left[[\hat{\bmy}_{\revision{\mathcal{I}}}]_b-\frac{\tau[\hat{\bmy}_{\mathcal{I}}]_b }{\sqrt{[\hat{\bmy}_{\mathcal{R}}]_b^2+[\hat{\bmy}_{\mathcal{I}}]_b^2}}\right]\nonumber \\ 
= &\, \textstyle 1-\tau \frac{[\hat{\bmy}_{\mathcal{R}}]_b^2}{{([\hat{\bmy}_{\mathcal{R}}]_b^2+[\hat{\bmy}_{\mathcal{I}}]_b^2)}^{3/2}}.
\end{align}
Note that  the derivative of $[\eta(\yhat,\tau)]_b$ has a discontinuity at $\tau=|\yhate{b}|$ (see \fref{fig:eta}) and thus, SURE is not defined for this value. Using \fref{eq:mu_0}, \fref{eq:mu_r} and \fref{eq:mu_i}, the complex-valued SURE expression \fref{eq:complexSURE} reduces to
\begin{align}
\textstyle \textit{SURE}_\tau = 
&\, \textstyle \revision{\textstyle\frac{1}{B}}\sum_{b=1}^{B}{\min\{|\hat{y}_b|,\tau\}}^2 +\Eo \nonumber \\ 
&\, \textstyle +\revision{\textstyle\frac{\Eo}{B}} \sum_{b:|\hat{y}_b|>\tau}\left(2-\tau \frac{1}{\sqrt{[\hat{\bmy}_{\mathcal{R}}]_b^2+[\hat{\bmy}_{\mathcal{I}}]_b^2}}-2\right)  \nonumber \\
&\, \textstyle + \revision{\textstyle\frac{\Eo}{B}} \sum_{b:|\hat{y}_b|<\tau}\left(0-2\right)\\
= &\, \textstyle \revision{\textstyle\frac{1}{B}}\sum_{b:|\hat{y}_b|<\tau}|\hat{y}_b|^2 + {\textstyle\frac{1}{B}}\sum_{b:|\hat{y}_b|>\tau}\tau^2 + \Eo \nonumber \\ 
&\, - \textstyle \revision{\textstyle\frac{\Eo}{B}}\tau\sum_{b:|\hat{y}_b|>\tau} \frac{1}{|\hat{y}_b|} - \revision{\textstyle\frac{2\Eo}{B}}\sum_{b:|\hat{y}_b|<\tau}1.
\end{align}

%% file: 8-SUREconvergence.tex
\section{Proof of \fref{thm:SUREconvergence}}
\label{app:SUREconvergence}
We now prove the convergence of SURE in \fref{eq:SUREconvergence}. 
In \cite[Lem. 4.14]{mousavi15}, the authors prove convergence of SURE to MSE in the real domain for the soft-thresholding function. We follow the same procedure for the complex domain. 
Using \cite[Thm. \uppercase\expandafter{\romannumeral 3}.15 \& \uppercase\expandafter{\romannumeral 3}.16]{maleki13}, we have that for any pseudo-Lipschitz function {$\gamma: \complexset^2 \to \reals$} the following equality holds:
\begin{align}\label{eq:lipschitz_stuff}
& \textstyle \lim\limits_{B \to \infty} \frac{1}{B} \sum_{b=1}^{B}\gamma(\eta(\hat{y}_b,\tau),\hat{h}_b) \notag \\ 
& \textstyle \qquad \qquad =  \Ex{}{\gamma(\eta(H+\sqrt{\Eo}Z,\tau),H)}\!.
\end{align}
{Here, $Z \sim \setC \setN(0,1)$ and $H$ is a random variable with the sparse distribution of a channel coefficient in the beamspace domain $\hat{h}_b$.} Using \fref{eq:lipschitz_stuff}, we have the following result
\begin{align} \label{eq:lipshcitz_entrywise}
& \textstyle \lim\limits_{B \to \infty} \frac{1}{B} \sum_{b=1}^{B} |\eta(\hat{y}_b,\tau)-\hat{y}_b|^2 \notag \\
& \qquad \qquad  = \Ex{\hat{y}_{\tilde{b}}}{|\eta(\hat{y}_{\tilde{b}},\tau)-\hat{y}_{\tilde{b}}|^2}\!,
\end{align}
where, $\hat{y}_{\tilde{b}}$ is any element of the random vector $\yhat$. The expression above can be rewritten as 
\begin{align}\label{eq:term1}
\textstyle \lim\limits_{B \to \infty} \frac{1}{B} \|\eta(\hat{\bmy},\tau)-\hat{\bmy}\|_2^2 = \Ex{\yhat}{\frac{1}{B} \|\eta(\yhat,\tau)-\yhat\|_2^2}\!.
\end{align}
Now, since $\frac{\partial {[\eta_{\mathcal{R}}(\hat{\bmy},\tau)]_b}}{\partial [\hat{\bmy}_{\mathcal{R}}]_b}+\frac{\partial {[\eta_{\mathcal{I}}(\hat{\bmy},\tau)]_b}}{\partial [\hat{\bmy}_{\mathcal{I}}]_b}$ is bounded, it is pseudo-Lipschitz. Hence, we can use \fref{eq:lipschitz_stuff} to obtain the following convergence result:
\begin{align}\label{eq:term3}\nonumber
& \textstyle \lim\limits_{B \to \infty} \frac{1}{B} \sum_{b=1}^{B} \left( \frac{\partial {[\mu_{\mathcal{R}}(\yhat)]_b}}{\partial [\yhat_{\mathcal{R}}]_b}+\frac{\partial {[\mu_{\mathcal{I}}(\yhat)]_b}}{\partial [\yhat_{\mathcal{I}}]_b}-2\right) \\
& \textstyle\quad = \frac{1}{B} \Ex{}{ \sum_{b=1}^{B} \left( \frac{\partial {[\mu_{\mathcal{R}}(\yhat)]_b}}{\partial [\yhat_{\mathcal{R}}]_b}+\frac{\partial {[\mu_{\mathcal{I}}(\yhat)]_b}}{\partial [\yhat_{\mathcal{I}}]_b}-2\right)}\!.
\end{align}
By summing \fref{eq:term1} and \fref{eq:term3}, combined with the fact that $ \frac{1}{B}\Ex{}{\|\yhat-\hhat\|_2^2} = \Eo$, {we have established that $\lim_{B \to \infty} \textit{SURE}_\tau = \Ex{}{\textit{SURE}_\tau}$. Finally, using \fref{thm:MSEappox}, we also prove that $\lim_{B \to \infty} \textit{SURE}_\tau = \textit{MSE}$.}